\documentclass[final,authoryear,5p,times,twocolumn]{elsarticle}

\usepackage{graphicx}

\usepackage{amssymb}

\usepackage{hyperref}
\usepackage{breakurl}

\usepackage{amsmath} 

\usepackage{listings} 
\usepackage{color}
\definecolor{lightgray}{rgb}{0.9,0.9,0.9}
\definecolor{darkgreen}{rgb}{0,0.4,0}
\lstdefinelanguage{vastmarkup}
{alsoletter={-,\.},
 morekeywords={vast,find_candidates,save.sh,load.sh,select_star_on_reference_image,lc,magnitude_calibration.sh,sextract_single_image,sysrem2,tiff2fits,modhead,listhead,identify.sh,three_column_ascii2vast.sh,hjd_input_in_TT,hjd_input_in_UTC,cute_lc,split_multiextension_fits},
 morekeywords={[2]-a,--selectbestaperture,-P,--UTC,V,-i,-x7,-ukf,-uf,-h,--help,-9,--ds9,-f,--nofind,-d,--debug,-p,--poly,-o,--photocurve,-P,--PSF,-r,--norotation,-e,--failsafe,-u,--UTC,-k,--nojdkeyword,-a5.0,--aperture5.0,-b200,--matchstarnumber200,-y3,--sysrem3,-x2,--maxsextractorflag2,-j,--position_dependent_correction,-J,--no_position_dependent_correction,-g,--guess_saturation_limit,-G,--no_guess_saturation_limit,-1,--magsizefilter,-2,--nomagsizefilter,-3,-4,--noerrorsrescale,-5,--starmatchraius,-6,--notremovebadimages,-a10,-up},
 emph={prompt},
 moredelim=**[is][\itshape]{@}{@},
 moredelim=**[is][\normalsize\bfseries]{~}{~},
    sensitive=true,
}

\lstdefinelanguage{novastmarkup}
{alsoletter={-,\.},
 morekeywords={cd,wget,tar,make},
 morekeywords={[2]-xvf},
 emph={prompt},
    sensitive=true,
}

\bibpunct{(}{)}{,}{a}{}{,} 

\journal{Astronomy \& Computing}

\begin{document}


\begin{frontmatter}



\title{{\scshape VaST}: a variability search toolkit$^\star$}


\author[1,2,3]{Kirill~V.~Sokolovsky}
\address[1]{IAASARS, National Observatory of Athens, Vas.~Pavlou \& I.~Metaxa, 15236~Penteli, Greece; email: kirx@noa.gr}
\address[2]{Sternberg Astronomical Institute, Moscow State University,
Universitetskii~pr.~13, 119992~Moscow, Russia}
\address[3]{Astro Space Center, Lebedev Physical Institute, Russian
Academy of Sciences, Profsoyuznaya~St.~84/32, 117997 Moscow, Russia}
\author[4]{Alexandr~A.~Lebedev}
\address[4]{Yandex School of Data Analysis,
Timura Frunze St.~11/2, 119021 Moscow, Russia}

\begin{abstract}
{\scshape Variability Search Toolkit (VaST)} is a software package designed to 
find variable objects in a series of sky images.
It can be run from a script or interactively using its graphical interface.
{\scshape VaST} relies on source list matching as opposed to image subtraction.
{\scshape SExtractor} is used to generate source lists and perform aperture or
PSF-fitting photometry (with {\scshape PSFEx}).
Variability indices that characterize scatter and smoothness of a
lightcurve are computed for all objects. Candidate variables are identified as objects
having high variability index values compared to other objects of similar brightness.
The two distinguishing features of {\scshape VaST} are its ability to
perform accurate aperture photometry of images obtained with non-linear detectors and handle complex image distortions.
The software has been successfully applied to images obtained with telescopes ranging from
0.08 to 2.5\,m in diameter equipped with a variety of detectors
including CCD, CMOS, MIC and photographic plates.
About 1800 variable stars have been discovered with {\scshape VaST}. 
It is used as a transient detection engine in the New Milky Way (NMW) nova patrol.
The code is written in {\scshape C} and can be easily compiled on the majority of {\scshape UNIX}-like systems.
{\scshape VaST} is free software available at \url{http://scan.sai.msu.ru/vast/}
\end{abstract}


\end{frontmatter}


\section{Introduction}
\label{sec:intro}

\renewcommand{\thefootnote}{$\star$}
\setcounter{footnote}{1}
\footnotetext{This code is registered at the ASCL with the code entry ascl:1704.005.}
\setcounter{footnote}{0}
\renewcommand{\thefootnote}{\arabic{footnote}}

Variable stars are important tracers of stellar evolution
\citep[e.g.][]{1975IAUS...67.....S}, fundamental stellar parameters
\citep{2010A&ARv..18...67T}, 3D structure of our Galaxy
\citep{2015ApJ...811..113P,2015ApJ...812L..29D,2016A&A...591A.145G} and
beyond \citep{2005MNRAS.359L..42L,2012ApJ...744..128S,2015AJ....149..183H,2017AcA....67....1J} 
as well as various astrophysical processes related to accretion
\citep{1996PASP..108...39O,2017PASP..129f2001M}, ejection \citep{2016MNRAS.460.3720R} and strong
magnetic fields \citep{1989MNRAS.236P..29C}.
It is believed that only a few per~cent of the variable stars easily accessible 
to ground-based photometry are currently known \citep{2015HiA....16..687S}.
The reason is that contemporary CCDs are very sensitive and small 
and hence image only a small field of view to a high limiting magnitude.

The next generation surveys 
Gaia \citep{2016A&A...595A...1G,2016A&A...595A.133C}, 
VVV \citep{2010NewA...15..433M}, 
Pan-STARRS \citep{2016arXiv161205560C},
LSST \citep{2014ApJ...796...53R},
NGTS \citep{2017arXiv171011100W}
and TESS \citep{2015JATIS...1a4003R}
employing large mosaic cameras (or multiple small cameras in case of NGTS and TESS)
are expected to greatly increase the number of known variable stars. 
Still, these surveys have their limitations in terms of observing cadence,
sky coverage, accessible magnitude range and survey lifetime. 
All this leaves room for variability searches with other instruments and different
observing strategies.
Photometric measurements needed to detect stellar variability are relatively
easy to perform (compared to spectroscopy and polarimetry of objects with
the same brightness), so even small-aperture telescopes are useful for
finding and studying variable stars.

Large time-domain surveys employ custom-built pipelines 
\citep[e.g.][]{2002ASPC..281..228B,2014PASP..126..674L,2015AJ....150..172K,2015ASPC..491...83M}
to perform photometric data reduction and variable object detection. 
These pipelines are fine-tuned for the particular 
equipment and observing strategies employed by these surveys and, while often
sharing many common pieces of code, require intervention of a software
engineer to adopt them to another telescope or camera.
Developing a purpose-built pipeline for a small observing project is often impractical. 
Instead, one would like to have data reduction software applicable to a variety of 
telescopes and cameras.

The problem of extracting variability information from surveys,
in practice, has not been completely solved. New variable stars are still being
identified in the NSVS survey data \citep[e.g.][]{2013PZP....13...16K,2014PZP....14....4S} 
while its observations were completed in 1999--2000 \citep{2004AJ....127.2436W}.
A number of recent ground-based exoplanet transit surveys, despite 
having sufficient photometric accuracy and sky coverage for detecting 
{\it the majority} of bright variable stars,
have so far provided only limited information on individual objects
\citep[e.g.][]{2013AJ....146..112R,2016A&A...587A..54N} or specific classes of objects
\citep[e.g.][]{2008AJ....135..850D,2011A&A...528A..90N,2012MNRAS.419..330M,2014MNRAS.439.2078H,2016arXiv160908449L}.
Better variability detection algorithms, open data-sharing policies and
interfaces to published time-series that allow non-trivial searches 
in the whole database rather than providing access to
a limited number of objects at a time are needed to fully exploit 
the information hidden in the data.
A software to perform variability search in a set of
lightcurves imported from a survey archive and visualize the search results 
may be useful for information extraction and debugging fully-automated search procedures.

Photographic plates used to be the primary type of light detectors in astronomy
in the 20th century. Direct images of the sky recorded on glass plates contain
information about the positions and brightness that celestial objects had decades
ago. This information may be useful on its own or as the first-epoch for 
comparison with modern CCD measurements.
The plates are stored in archives in observatories around the world.
Many observatories are digitizing their collections in an
effort to preserve the information stored on the plates and make it more accessible.
At the time of writing, only the DASCH\footnote{\url{http://dasch.rc.fas.harvard.edu/}}
\citep{2012IAUS..285...29G}
and 
APPLAUSE\footnote{\url{https://www.plate-archive.org}}
\citep{2014aspl.conf...53G,2014aspl.conf..127T}
archives provide source catalogs and photometry derived from the plates
while
others\footnote{\url{http://dc.zah.uni-heidelberg.de/lswscans/res/positions/q/info}}$^,$\footnote{\url{http://plate-archive.hs.uni-hamburg.de/index.php/en/}} provide only images.
Performing photometry on digitized photographic images is a non-trivial task
\citep{2005MNRAS.362..542B,2013PASP..125..857T,2016arXiv160700312W}
that cannot be done well with conventional photometry software
developed for CCD images.
The conventional software relies on the assumption that an image sensor responds 
linearly to the number of incoming photons. This assumption is violated for photographic 
emulsion as well as for some types of contemporary light detectors including
microchannel plate intensified~CCDs (MICs) used in space-based UV-sensitive
telescopes Swift/UVOT (\citealt{2008MNRAS.383..627P,2010MNRAS.406.1687B},
see also \citealt{2014Ap&SS.354...89B}),
XMM/OM \citep{2001A&A...365L..36M} and fast ground-based cameras
\citep{2012ASInC...7..219K}.
There is a need for a user-level software capable of performing photometry on images 
obtained with non-linear detectors.

A number of software packages for detection of variable objects have been
developed recently. Some of them feature a graphical user interface (GUI)
and are aimed at processing small datasets, while others have command-line
interfaces and are meant as a complete data processing pipeline or as building blocks for constructing one.
{\scshape LEMON}\footnote{\url{https://github.com/vterron/lemon}}
is a pipeline based on {\scshape SExtractor}\footnote{\url{http://www.astromatic.net/software/sextractor}}
\citep{1996A&AS..117..393B} and {\scshape PyRAF}
for automated time-series reduction and analysis
\citep{2011hsa6.conf..755T}. It takes a series of \texttt{FITS} 
images as an input and constructs lightcurves of all objects.
{\scshape PP}\footnote{\url{https://github.com/mommermi/photometrypipeline}} \citep{2017A&C....18...47M} is
an automated {\scshape Python} pipeline based on {\scshape SExtractor} and {\scshape SCAMP}
\citep{2006ASPC..351..112B}. {\scshape PP} produces calibrated photometry from 
imaging data with minimal human interaction. While originally designed for asteroid
work, the code is be applicable for stellar photometry. 
{\scshape LEMON} and {\scshape PP} are very similar to {\scshape VaST} in spirit, 
while they differ in technical implementation and user interface.
{\scshape C-Munipack/Muniwin}\footnote{\url{http://c-munipack.sourceforge.net/}}
offers a complete solution for reducing observations of variable
stars obtained with a CCD or DSLR camera. It runs on Windows and Linux
and provides an intuitive GUI. 
{\scshape FotoDif}\footnote{\url{http://www.astrosurf.com/orodeno/fotodif/}} is a
Windows CCD photometry package providing capabilities similar to Muniwin.
{\scshape Astrokit}\footnote{\url{http://astro.ins.urfu.ru/en/node/1330}}
\citep{2014AstBu..69..368B,2016MNRAS.461.3854B} corrects for atmospheric 
transparency variations by selecting an
optimal set of comparison stars for each object in the field. 
{\scshape IRAF}'s\footnote{\url{http://iraf.noao.edu/}}
\citep{1986SPIE..627..733T}
{\scshape PHOT/APPHOT} task is meant to be used to generate input photometric data for
{\scshape Astrokit}. Variable star candidates are selected in {\scshape Astrokit} with 
Robust Median Statistics \citep{2007AJ....134.2067R}.
{\scshape FITSH}\footnote{\url{https://fitsh.net/}} \citep{2012MNRAS.421.1825P}
is a collection of tasks for advanced image manipulation (including stacking)
and lightcurve extraction using aperture, image subtraction, analytic profile modeling
and PSF-fitting photometry.
{\scshape VARTOOLS}\footnote{\url{http://www.astro.princeton.edu/~jhartman/vartools.html}}
\citep{2016A&C....17....1H} implements in C a collection of advanced
lightcurve analysis methods providing a command-line interface to them.
{\scshape HOTPANTS}\footnote{\url{http://www.astro.washington.edu/users/becker/v2.0/hotpants.html}}
\citep{2015ascl.soft04004B}
is designed to photometrically align one input image with another, after
they have been astrometrically aligned with software like 
{\scshape WCSremap}\footnote{\url{http://www.astro.washington.edu/users/becker/v2.0/wcsremap.html}},
{\scshape SWarp} \citep{2002ASPC..281..228B} or {\scshape Montage} \citep{2010arXiv1005.4454J}. 
{\scshape HOTPANTS} is an implementation of the \cite{2000A&AS..144..363A} algorithm for image subtraction. 
The program is intended as a part of a transient detection/photometry pipeline. 
{\scshape ISIS}\footnote{\url{http://www2.iap.fr/users/alard/package.html}}
\citep{1998ApJ...503..325A} is a complete package to process CCD images using the image
subtraction method. It finds variable objects in the subtracted
images and builds their light curves from a series of CCD images. 
{\scshape DIAPL}\footnote{\url{http://users.camk.edu.pl/pych/DIAPL/index.html}}
\citep{2017arXiv170909572R} is able to identify variable stars via image subtraction, implemented in C.
{\scshape TraP}\footnote{\url{https://github.com/transientskp/tkp}}
\citep{2015A&C....11...25S} is a {\scshape Python} and {\scshape SQL} based
pipeline for detecting transient and variable sources in a stream of astronomical images. 
It primarily targets LOFAR radio astronomy data, but is also applicable to a
range of other instruments (including optical ones). 

Most of the above packages were not available at the time {\scshape VaST}
development was started \citep{2005ysc..conf...79S}. 
Many of them cannot construct lightcurves without finding
a plate solution with respect to an external star catalog. None of the above
software addresses the issue of photometry with non-linear imaging detectors.
{\scshape VaST} provides a combination of features not yet offered by other
software, including the ability to process thousands of images with
tens of thousands of stars and interactively display the results in a GUI.

{\scshape VaST} is designed as a user-friendly software implementing 
the full cycle of photometric reduction from calibrating images to producing
lightcurves of all objects within a field of view and detecting variable
ones. {\scshape VaST} is capable of handling images obtained with non-linear 
detectors such as photographic plates and MICs. 
The software may be applied to images obtained with telescopes of any size
with minimal configuration.
{\scshape VaST} can be used interactively 
to inspect a set of images of one field
in a PGPLOT\footnote{\url{http://www.astro.caltech.edu/~tjp/pgplot/}}-based GUI. 
As soon as optimal source extraction and lightcurve post-processing parameters
have been identified in the interactive mode, {\scshape VaST} 
may proceed non-interactively and produce lightcurves for subsequent variability 
searches with {\scshape VARTOOLS} or custom-built scripts.
{\scshape VaST} lightcurve visualization tools and built-in variability statistics
routines may also be used to inspect and process lightcurves obtained with other software.

The outline of the paper is as follows. 
Section~\ref{sec:designoveview} presents the overall design of the program
centered around the idea of distinguishing variable stars from non-variable
ones using ``variability indices''.
Section~\ref{sec:dataproc} describes the steps performed by {\scshape VaST} 
when processing a set of images.
Section~\ref{sec:transientssec} presents a specialized transient detection mode
that does not rely on variability indices.
In Section~\ref{sec:remarks} we present some general remarks
on the {\scshape VaST} development procedures. 
Section~\ref{sec:applications} 
gives an account of several notable examples of applying {\scshape VaST} to real data processing.
We present a summary in Section~\ref{sec:summ}.
\ref{sec:usecases} describes common use cases that can be solved with {\scshape VaST}. 
\ref{sec:vastcommandlineopt} presents the list of the command-line arguments to be
used with the main program of the toolkit.
\ref{sec:logfiles} describes the log files that summarize image processing results.

\section{{\scshape VaST} design and the challenge of variability detection}
\label{sec:designoveview}

{\scshape VaST} is designed to accomplish three main tasks:
\begin{enumerate}
\item Construct lightcurves of all objects visible at the input series of images of a given star field.
\item Compute variability indices to quantify ``how variable'' each lightcurve appears to be.
\item Visualize the variability indices, lightcurves and images to let the user decide which objects are actually variable.
\end{enumerate}
In addition, {\scshape VaST} includes tools to calibrate the magnitude
zero-point of the produced lightcurves, check if a given object is a
cataloged variable, apply the heliocentric correction to a lightcurve, etc.
The software modules implementing each of the above tasks are united by 
a common internal lightcurve file format. They can be run together or independently of each other.
All data processing is done within the {\scshape VaST} working directory
which contains all the necessary binaries and scripts, so the modules can locate each 
other without the need to set environment variables. This also allows for 
an easy ``unpack and compile'' installation procedure described in \ref{sec:compilingvast}.

The input to {\scshape VaST} may be either a set of images (for a practical example see \ref{sec:simpleccdsearch}) 
or a set of lightcurves constructed using other software or imported from an archive
(\ref{sec:importlightcurves}). 
The output is a set of lightcurves and variability indices computed for each lightcurve.
The indices and lightcurves can be either 
visualized and searched for variable objects interactively or
exported for further processing with external software.

The fundamental problem in optical variability detection is that the accuracy of
photometric measurements is not precisely known as the measurements may 
be affected by various systematic effects or corrupted. 
These effects include, but are not limited to: 
\begin{itemize}
 \item Residual sensitivity variations across the detector that
were not taken out by flat fielding. These may be both large- (vignetting,
defocused images of dust grains in optical path) and small-scale (individual
dead pixels). 
The object's image moves across the detector due to imperfect 
guiding resulting in apparent variations in brightness.
 \item Variation of atmospheric extinction across the field.
 \item Differential extinction in the atmosphere resulting in apparent change
(as a function of airmass) in relative brightness of stars having different colors. 
 \item Charge transfer inefficiency in a CCD \citep{2015MNRAS.453..561I}.
 \item Changing seeing conditions resulting in varying degree of overlap
between images of nearby objects.
 \item If the instrument's point spread function (PSF) is not point-symmetric
(for example it may have diffraction spikes)
the amount of blending between objects will change with changing
parallactic angle.
 \item The PSF shape varies across the image resulting in different fraction
of object's light falling within the aperture (in the case of aperture
photometry) or presenting a challenge of accurate PSF variations
reconstruction (in PSF-fitting photometry).
 \item Resampling and stacking images may also have the unintended
consequence of distorting brightness of some sources.
\end{itemize}

The solution adopted by the community is to rely on the assumption that no instrumental
effect can in practice mimic a reasonably well sampled lightcurve of a
variable star of a known type. 
While this is a reasonable assumption for periodic variables, 
for irregular ones it is necessary to have an idea of the practical limits of
the amplitude and timescale of instrumental effects that may be found in a given photometric dataset.

In order to distinguish candidate variables from noise in the absence of
reliable estimates of measurement errors one may assume that the majority of
stars are not variable at the few per~cent level of photometric accuracy
\citep{2008AN....329..259H,2014aspl.conf...79S} 
typically achieved in ground-based survey observations.
Using this assumption one may select as candidate variables the objects that
appear ``more variable'' than the majority of objects in this dataset.
Among these candidates there will be true variable objects as well as false
candidates whose measurements were severely corrupted by some rare
instrumental effect like an object's image falling on a dead pixel. 
If the instrumental effect is not rare, e.g. if bad pixels are abundant 
and affect measurements of many stars, the affected stars will not stand out
as ``more variable'' compared to the other stars in the dataset.
It is ultimately up to a human expert to investigate lightcurves and images of 
the candidate variables and judge if the measurements of a particular object 
are trustworthy or have obvious problems (bad pixels, blending, etc) 
making them unreliable.

There is a number of ways to characterize a ``degree of variability'' in a lightcurve. 
We call the variability indices all the values that:
\begin{itemize}
 \item Quantify the scatter of brightness measurements: $\sigma$, 
MAD, 
IQR, 
 etc. \citep{2017A&A...604A.121F,2017MNRAS.464..274S}.
 \item Quantify the smoothness of the lightcurve like
the $I$ index \citep{1993AJ....105.1813W} or $1/\eta$ \citep{vonneumann1941}.
 \item Sensitive to both smoothness and scatter
(Stetson's $J$ index), maybe also taking into account the shape of the measured brightness distribution
\citep[e.g. $L$ index][]{1996PASP..108..851S,2016A&A...586A..36F}.
 \item Characterize the strength of a periodic signal in the lightcurve
\citep[e.g.][]{2012AJ....143..140F,2016arXiv160503571S}.
\end{itemize}
In the absence of detailed information about measurement errors and
instrumental effects one should consider the above parameters
computed for a given lightcurve in the context of other lightcurves in this
dataset to decide if a given variability index value corresponds to a
variable object or not. 
The indices are not equally sensitive to all types of variability.
Some of them are more susceptible to individual outliers (corrupted
measurements) than the others.
Scatter-based indices are generally sensitive to variability of any kind while the
indices that characterize smoothness are more sensitive than scatter-based
indices to objects that vary slowly compared to the typical observing
cadence. That comes at the cost of these indices being insensitive to
variability on timescales shorter than the observing cadence.
Table~\ref{tab:varindex} presents the list of variability indices computed by {\scshape VaST}.
We refer to \cite{2017MNRAS.464..274S} for a detailed discussion and
comparison of these indices.

In the following Section we present a detailed description of all the 
processing steps from reading input images to constructing lightcurves, 
computing and visualizing the variability indices.
\ref{sec:usecases} gives practical examples of using the code.

\begin{table}[ht!]
    \caption{Variability indices computed by {\scshape VaST}.}
    \label{tab:varindex}
    \begin{tabular}{r@{~~~}l}
    \hline\hline
Index                                               & Reference  \\
    \hline
\multicolumn{2}{c}{~} \\
\multicolumn{2}{c}{\it Indices quantifying lightcurve scatter} \\
weighted std. deviation -- $\sigma$             & \text{\cite{2008AcA....58..279K}}  \\
clipped $\sigma$ -- $\sigma_{\rm clip}$             & \text{\cite{2008AcA....58..279K}}  \\
median abs. deviation -- ${\rm MAD}$                & \text{\cite{2016PASP..128c5001Z}}  \\
interquartile range -- ${\rm IQR}$                  & \text{\cite{2017MNRAS.464..274S}}  \\
reduced $\chi^2$ statistic -- $\chi_{\rm red}^2$    & \text{\cite{2010AJ....139.1269D}}  \\
robust median stat. -- ${\rm RoMS}$             & \text{\cite{2007AJ....134.2067R}}  \\
norm. excess variance -- $\sigma_{\rm NXS}^2$       & \text{\cite{1997ApJ...476...70N}}  \\
norm. peak-to-peak amp. -- $v$                      & \text{\cite{2009AN....330..199S}}  \\
\multicolumn{2}{c}{~} \\
\multicolumn{2}{c}{\it Indices quantifying lightcurve smoothness} \\
autocorrelation -- $l_1$                            & \text{\cite{2011ASPC..442..447K}}  \\
inv. von~Neumann ratio -- $1/\eta$                  & \text{\cite{2009MNRAS.400.1897S}}  \\
Welch--Stetson index -- $I_{\rm WS}$                 & \text{\cite{1993AJ....105.1813W}}  \\
flux-independent index -- $I_{\rm fi}$              & \text{\cite{2015A&A...573A.100F}}  \\
Stetson's~$J$ index                                 & \text{\cite{1996PASP..108..851S}}  \\
time-weighted Stetson's~$J_{\rm time}$              & \text{\cite{2012AJ....143..140F}}  \\
clipped Stetson's~$J_{\rm clip}$                    & \text{\cite{2017MNRAS.464..274S}}  \\
Stetson's~$L$ index                                 & \text{\cite{1996PASP..108..851S}}  \\
time-weighted Stetson's~$L_{\rm time}$              & \text{\cite{2012AJ....143..140F}}  \\
clipped Stetson's~$L_{\rm clip}$                    & \text{\cite{2017MNRAS.464..274S}}  \\
consec. same-sign dev. -- {\it Con.}                & \text{\cite{2000AcA....50..421W}}  \\
$S_B$ statistic                                     & \text{\cite{2013A&A...556A..20F}}  \\
excursions -- $E_x$                                 & \text{\cite{2014ApJS..211....3P}}  \\
excess Abbe value -- $\mathcal{E}_\mathcal{A}$      & \text{\cite{2014A&A...568A..78M}}  \\
\multicolumn{2}{c}{~} \\
\multicolumn{2}{c}{\it Indices quantifying magnitude distribution shape} \\
Stetson's~$K$ index                                 & \text{\cite{1996PASP..108..851S}}  \\
kurtosis                                            & \text{\cite{1997ESASP.402..441F}}  \\
skewness                                            & \text{\cite{1997ESASP.402..441F}}  \\
    \hline
    \end{tabular}
\end{table}

\section{Data processing flow}
\label{sec:dataproc}

The primary input for {\scshape VaST} is a series of \texttt{FITS}\footnote{\url{http://fits.gsfc.nasa.gov/}} images.
The images are expected to be taken with the same telescope-camera-filter
combination and cover the same area of the sky. The images may be shifted,
rotated and reflected with respect to each other, but should overlap by at least
40\,per~cent
with the sky area covered by the first image.
The first image supplied to {\scshape VaST} serves as the astrometric and photometric reference.
If the input images vary in quality, it is up to the user to select a good
reference image and put it first on the command line (\ref{sec:refimage}).
Below we describe the main processing steps performed to extract lightcurves from a set of images.

\subsection{Reading metadata from the \texttt{FITS} header}
\label{sec:fitsheader}

Information about the observing time, image dimensions 
and CCD gain information (that is needed to accurately
compute contribution of the Poisson/photon noise to the total photometric error), 
is extracted from \texttt{FITS} header of each input image. 
The observing time assigned to each image is the middle of the exposure. 
The exposure start time is derived from one or two (if date and
time are given separately) keywords from the following list:
\texttt{DATE-OBS},
\texttt{TIME-OBS},
\texttt{EXPSTART},
\texttt{UT-START},
\texttt{START}.
The exposure time in seconds is extracted from \texttt{EXPOSURE} or
\texttt{EXPTIME}. If none of the two keywords are present in the header, 
the difference between the observation start and its middle-point is assumed
to be negligible (the exposure time is set to 0). The derived time is
converted to Julian date (JD).

If no keyword describing the exposure start time is found, but 
the \texttt{JD} keyword is present in the header, the keyword is
assumed to correspond to the middle of exposure. 
This simplified convention for recording the observation time in 
the \texttt{FITS} header 
has been introduced within
the program of digitizing photographic plates of 
Sternberg Astronomical Institute's collection
\citep{2008AcA....58..279K,2010ARep...54.1000K,2014aspl.conf...79S}.
This is different from the definition of the \texttt{JD} keyword adopted by 
the widely-used CCD camera control software {\scshape MaxIm~DL}. However,
{\scshape MaxIm~DL} will always write the \texttt{DATE-OBS} keyword
which, if present, is used by {\scshape VaST} instead of \texttt{JD}.
If {\scshape VaST} fails to recognize the format in which the observing time
is specified in the FITS header of your images, please send an example image to the
authors.

{\scshape VaST} supports the following two time systems: 
Coordinated Universal Time (UTC) 
and 
Terrestrial Time (TT). 
The Earth rotation period is a few milliseconds longer than
86400~SI~seconds and is changing irregularly when measured at this level of accuracy.
To keep UTC timescale in sync with the Earth rotation, 
a leap second 
is introduced every few years.
The UTC time, being the basis of civil time, is readily available in
practice through GPS receivers, {\scshape NTP} servers and radio broadcast, so the
time of most images is recorded in UTC.
The observing time from some instruments, notably Swift/UVOT, is recorded in TT.
The advantage of TT is that this timescale is continuous, not interrupted by leap seconds.
At each moment of time, the two timescales are related as
$${\rm TT} = {\rm UTC} + ({\rm TAI} - {\rm UTC}) + 32.184\,{\rm sec}$$
where $({\rm TAI} - {\rm UTC})$ is the difference between the 
International Atomic Time (TAI) and UTC for 
a given date\footnote{{\scshape VaST} will automatically get the current $({\rm TAI} - {\rm UTC})$
values from \url{http://maia.usno.navy.mil/ser7/tai-utc.dat}}.
As of spring 2017, the difference between TT and UTC is 69.184\,sec.

The program tries to determine the time system in which the
observing time is expressed in the \texttt{FITS} header by searching for 
the \texttt{TIMESYS} key or parsing the
comment field for the keyword used to get the time. If neither UTC nor TT are
explicitly mentioned in the header, the input time is assumed to be
UTC. {\it The observing time in the lightcurves created by {\scshape VaST} is
by default expressed in terrestrial time -- JD(TT).} The conversion from UTC to TT
may be disabled (\ref{sec:vastcommandlineopt}).

No heliocentric correction is applied to the lightcurves by default, as by
design of the program, it is possible to produce lightcurves without knowing
the absolute celestial coordinates of the imaged objects.
If the coordinates are known, the correction may be applied after constructing a
lightcurve as described in \ref{sec:heliocorr}.
See \cite{2010PASP..122..935E} for a detailed discussion of accurate timing
problems in the context of optical photometric observations.

\subsection{Source extraction and photometry}
\label{sec:sextraction}

{\scshape VaST} relies on {\scshape SExtractor} for detecting sources, 
measuring their positions (in pixel coordinates) and brightness.
The source extraction parameters are set in the {\scshape SExtractor}
configuration file \texttt{default.sex} located in the {\scshape VaST}
directory.
Depending on the command-line settings (\ref{sec:vastcommandlineopt}), more than one run of {\scshape SExtractor} may be
needed for each image: 

{\it option~I)}~If all images are to be measured with a circular aperture
of the fixed diameter, only a single {\scshape SExtractor} run is performed
for each image.

{\it option~II)}~If the circular aperture diameter is allowed to vary 
between images to account for seeing changes \citep[e.g.][]{2017AJ....153...77C},
a preliminary {\scshape SExtractor} run is performed to determine the
typical source size. 
For each source the maximum spatial r.m.s. dispersion of its profile 
is computed ({\scshape SExtractor} output parameter \texttt{A\_IMAGE}).
The aperture diameter for each image is computed as 
$C \times {\rm median(\texttt{A\_IMAGE})}$, where the median is computed over all the
detected sources passing selection criteria 
and the default value of $C=6.0$.

{\it option~III)}~If the program is running in the PSF-fitting photometry
mode, multiple {\scshape SExtractor} passes over the input image are
required. First, a preliminary run is performed to determine seeing, 
as in the {\it option~II)} above. Then {\scshape SExtractor} is run with
conservative detection limits to extract bright sources and save their
small images ({\scshape SExtractor} output parameter \texttt{VIGNET}).
These small images are used by 
{\scshape PSFEx}\footnote{\url{http://www.astromatic.net/software/psfex}}
\citep{2011ASPC..442..435B} to construct a model of spatially-variable point
spread function (PSF). The size of these small images as well as limits on
size of objects accepted for PSF reconstruction are determined based on
seeing. The PSF reconstruction step is followed by PSF-fitting with {\scshape SExtractor}.
Objects that provide a bad fit (non-stellar, heavily blended, corrupted or
overexposed objects) are discarded from the output catalog.
Finally, an aperture photometry run is conducted with {\scshape SExtractor}
(again, same as in {\it option~II}). This step is performed for technical
reasons: a catalog that includes all the objects, even the overexposed
ones, is needed at later processing steps to perform absolute astrometric calibration discussed
in Section~\ref{sec:astrometry}.

The input images are expected to be well-focused. {\scshape VaST}
may be used to process strongly-defocused images with ``donut-shaped stars'' given 
a careful selection of the processing parameters including source detection
and deblending thresholds (\ref{sec:finetuningsourceextraction}) and
manually specified aperture size and star matching radius
(Section~\ref{sec:match}, \ref{sec:vastcommandlineopt}).

{The list of sources detected by {\scshape SExtractor} may be filtered
to remove saturated, blended and extended sources as their brightness measurements 
are less accurate and such sources may appear as spurious candidate variables
later in the analysis. Saturated sources are removed based on the source
detection flag set by {\scshape SExtractor}.
The filtering of blended and extended sources may be accomplished using {\scshape SExtractor}
flags (if {\scshape SExtractor} deblending was successful) and by identifying outliers 
in the magnitude-size plot (Figure~\ref{fig:magsizefilter}).
The sources that were not properly deblended will appear larger than 
the other sources of similar brightness.
Rejection of blended sources is done using individual images rather than
using information about source size and detection flags collected from the full series of images
because a generally ``good'' source may appear blended on some images where
it is affected by a cosmic ray hit, a CCD cosmetic defect or a particularly
bad seeing. It is desirable to reject only measurements of the source
brightness associated with the affected images while keeping the good
measurements for further analysis.
The acceptable {\scshape SExtractor} flag values
(\ref{sec:finetuningsourceextraction}, \ref{sec:transients}) and the magnitude-size filter may
be controlled using {\scshape VaST} command line options (\ref{sec:vastcommandlineopt}).

\begin{figure}
 \centering
 \includegraphics[width=0.48\textwidth,clip=true,trim=0cm 0cm 0cm 0cm]{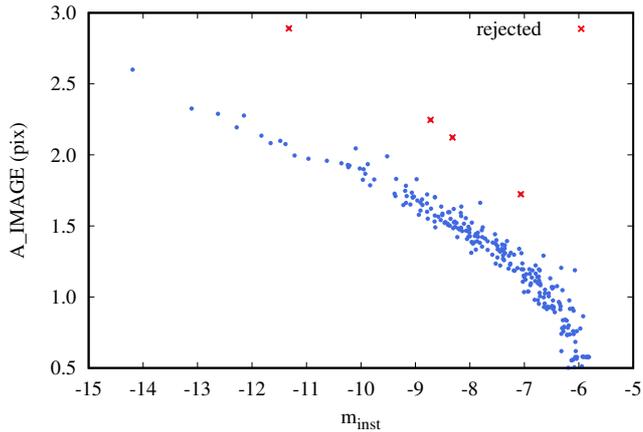}
 \caption{Semi-major axis length ({\scshape SExtractor} output parameter \texttt{A\_IMAGE})
 as a function of instrumental magnitude for sources detected on a CCD frame.
 Crosses mark sources rejected as being too large for their magnitude (blended or extended).}
 \label{fig:magsizefilter}
\end{figure}

\subsection{Cross-matching source lists}
\label{sec:match}

{\scshape VaST} does not require any world coordinate system (WCS;
\citealt{2002A&A...395.1061G}) information to be present in the headers of the
input \texttt{FITS} images. Instead, the code finds a linear transformation of pixel
coordinates of objects detected on each image to the pixel coordinate system
of the first image (taken as the astrometric reference).\footnote{The use of internal pixel-based coordinate
system instead of celestial coordinates may seem strange nowadays, 
but was natural back in 2004 when we started developing {\scshape VaST}.
Before the release of the {\scshape Astrometry.net} software
\citep{2010AJ....139.1782L,2008ASPC..394...27H} there was no easy way to
assign WCS to ``random'' images having no a~priori
information in \texttt{FITS} header about the image center, scale and orientation
(which was often the case for images obtained with non-robotic telescopes).}
The correct transformation is found by identifying similar triangles
constructed from the brightest stars detected on the current and reference images.
When we created this algorithm, we were unaware that analogous star lists matching algorithms based on finding 
similar triangles were proposed by other authors
\citep{1986AJ.....91.1244G,1995PASP..107.1119V,2006PASP..118.1474P,2007PASA...24..189T}\footnote{A
C implementation of \cite{2007PASA...24..189T} algorithm may be found at\\\url{http://spiff.rit.edu/match/}}.
A similar algorithm has also been independently proposed by 
P.~B.~Stetson\footnote{\url{https://ned.ipac.caltech.edu/level5/Stetson/Stetson5_2.html}}.
While sharing the basic idea that angles and side length ratios of a triangle are not changed by
translation, rotation, scale change and flip, the various implementations of
the algorithm differ in how the lists of triangles are constructed from the
input star lists, how the triangles are matched to find the initial guess of the
coordinate frame transformation and how this initial transformation is
further refined.

Our cross-matching algorithm goes through the following steps for the lists
of sources (assumed to be mostly stars) derived from each image:

\begin{enumerate}
\item Sort the detected stars in magnitude.
\item Select $N$ brightest stars having acceptable {\scshape SExtractor} flags as the reference. 
The value of $N$ is discussed below.
\item Construct a set of triangles from the selected $N$ stars using the
following two algorithms together:
\begin{enumerate}
\item For each reference star find the two nearest reference stars to form a
triangle.
\item For each reference star having the index $n$ in the magnitude-sorted
list, 
form triangles from stars having
indices: ($n$, $n+1$, $n+2$),  ($n$, $n+1$, $n+3$),  ($n$, $n+1$, $n+4$), 
($n$, $n+1$, $n+5$),  ($n$, $n+2$, $n+3$),  ($n$, $n+2$, $n+4$),  ($n$, $n+2$, $n+5$),  
($n$, $n+3$, $n+4$),  ($n$, $n+3$, $n+5$) and ($n$, $n+4$, $n+5$).
\end{enumerate}
In practice, a combination of these two algorithms allows one to obtain overlapping
lists of triangles for virtually any pair of reference star lists if these
lists overlap.
\item Similar triangles are identified in the lists 
by comparing ratios of the two smaller sides
to the larger side of each triangle in the first list 
(corresponding to the reference image)
to these ratios for triangles in the second list
(corresponding to the current image).
\item Accept only the similar triangles that have approximately the same scale to
improve stability of the algorithm.
\item For each pair of similar triangles construct a linear transformation 
between the pixel coordinates at the current frame and pixel coordinates 
at the reference frame. The transformations are applied to the $N$ reference
stars.
\item The transformation that allowed {\scshape VaST} to match the highest number of
reference stars is now applied to all the detected stars on the frame.
The residuals along the two axes (dX,dY) between the predicted and measured
positions of each matched star are recorded.
\item The residuals dX and dY are approximated as linear function of
coordinates (X,Y) on the reference frame by least-square fitting planes into
the 3D datasets (X,Y,dX) and (X,Y,dY). The derived least-square coefficients
are used to correct the coordinates of all sources computed from the initial
linear transformation determined from a pair of matched triangles. This
correction is necessary for large images matched using a pair of small
triangles (likely constructed by the algorithm (a) above). Since the
positions of stars forming the triangle are measured with a limited
accuracy, the initial linear transformation will be inaccurate for stars 
far away from the triangle.
\item Use the corrected positions to match stars to the ones detected on the
reference image.
A custom spatial indexing scheme is used to avoid the computationally expansive step of
calculating distances from each star in the current image list to each 
star in the reference list \citep[see the discussion by e.g.][]{2017PASP..129b4005R}.
The reference list used in this last step is complemented by stars detected on
previously matched images even if they were not detected 
(or rejected because of unacceptable flag values) on the reference image.
Only the stars originally detected on the reference image are used in the
reference list to construct triangles.
\end{enumerate}
The main input parameter of this algorithm is the number of reference stars
$N$. {\scshape VaST} is trying to match images probing various values of $N$ 
in the range 100--3000. The lower value is sufficient to match most
narrow-field images while the upper value is set by the requirement to
perform the match in a reasonably short time. 
The secondary input parameter is the matching radius -- the maximum
acceptable distance between the source position measured on the reference image 
and the one measured on the current image and transformed to the coordinates
system of the reference image. If the distance is larger than the matching radius
the source detected on the current image is assumed to be different from the
one detected on the reference image (no match). By default, the matching
radius is taken as $0.6$ times the aperture diameter for the current image
(it will vary with aperture from image to image depending on seeing).
Alternatively, the user may specify a fixed matching radius in pixels on the
{\scshape VaST} command line (\ref{sec:vastcommandlineopt}). 
If for a given source detected on the current image multiple sources are
found within the matching radius on the reference image, the nearest source 
is taken as the correct match.

If an image could not be matched because of its bad quality 
(e.g. many hot pixels/cosmic rays confused for real sources) 
or has insufficient overlap with the reference image, 
the unmatched image is discarded and {\scshape VaST} continues 
to the next image in the series. The matching results are logged
as described in \ref{sec:logfiles}.

The above image matching algorithm assumes a linear transformation
between images. This assumption may be approximately correct even for
highly-distorted wide-field images as long as the relative shifts between
the distorted images are small. For a successful match, the coordinate 
transformation error at image edges needs to be smaller than the matching 
radius and typical distance between stars (to avoid confusion). 
The requirements for star matching are less strict than they would be 
for image stacking (that demands sub-pixel coordinates transformation accuracy).
In practice, the matching algorithm works well for sky images obtained with 
lenses having a focal length of 100\,mm or longer.

\begin{figure*}
 \centering
 \includegraphics[width=0.48\textwidth,clip=true,trim=1.5cm 0cm 1.5cm 0cm]{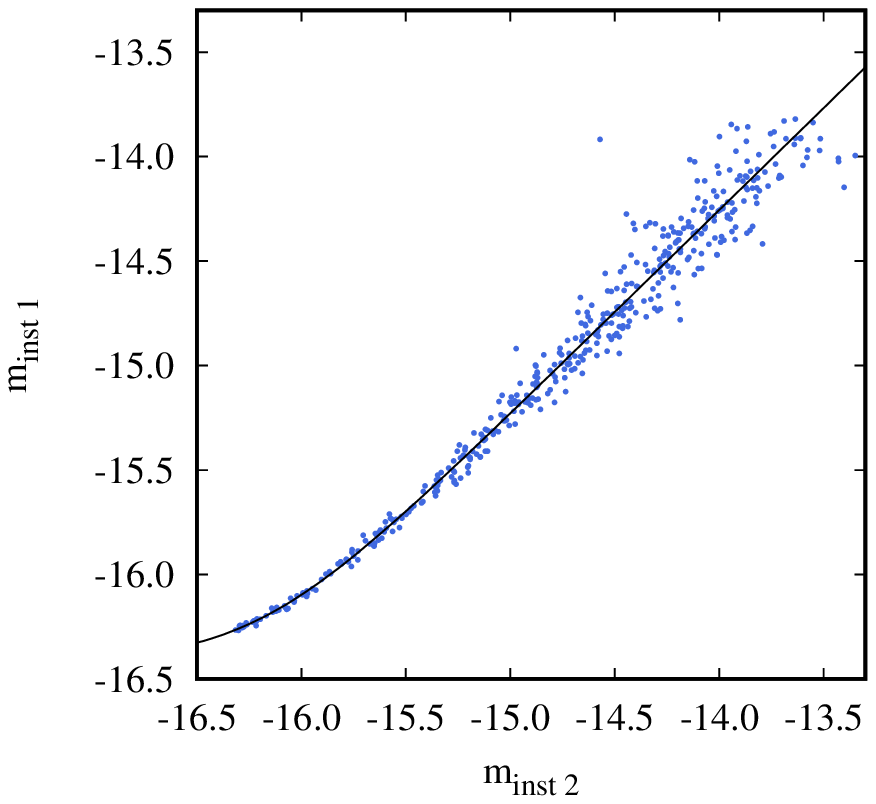}
 \includegraphics[width=0.48\textwidth,clip=true,trim=1.5cm 0cm 1.5cm 0cm]{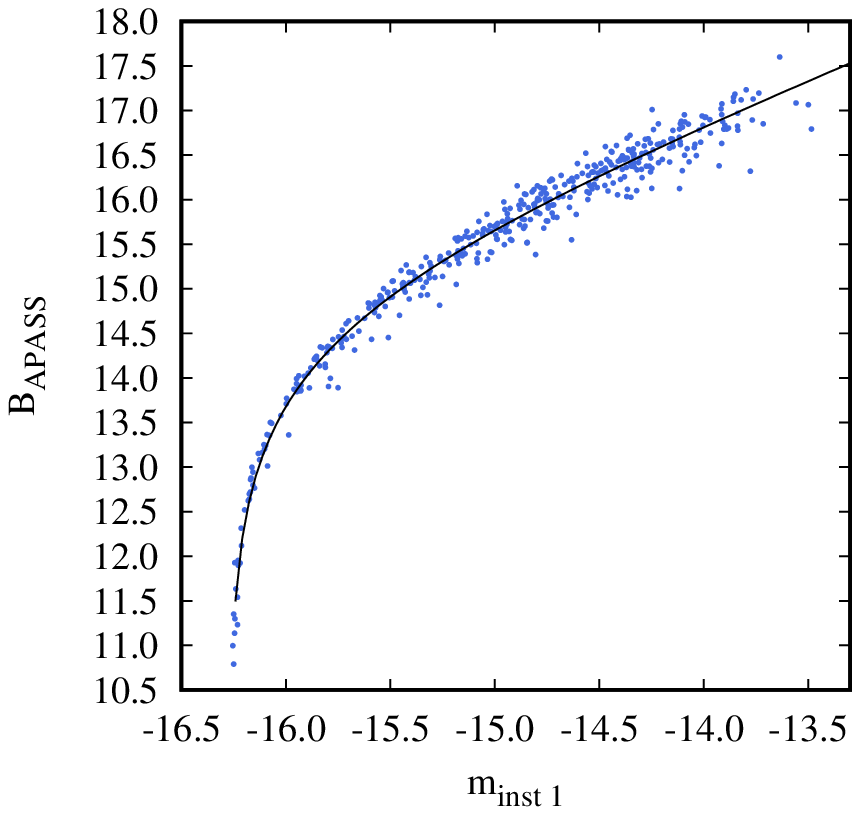}
 \caption{Two steps of magnitude calibration for aperture photometry using digitized photographic images. 
Left panel: measured instrumental magnitude on the reference frame, $m_{\rm inst~1}$, 
as a function of instrumental magnitude on the current frame (Section~\ref{sec:calibinstmag}). Right panel:
APASS $B$ magnitude as a function of mean instrumental magnitude in the reference
frame system (Section~\ref{sec:photometry}). The best-fit photocurve (shown in black) corresponds to
Equation~(\ref{eq:invphotocurve}) on the left panel and Equation~(\ref{eq:photocurve})
on the right.}
 \label{fig:magcalib}
\end{figure*}

\subsection{Cross-calibration of instrumental magnitudes}
\label{sec:calibinstmag}

The aperture and PSF-fitting magnitudes are measured by {\scshape SExtractor}
with respect to an arbitrary zero-point which differs from image to image.
{\scshape VaST} uses the stars matched with the reference image to convert 
magnitude scales of all images to the instrumental magnitude scale 
of the reference image.
While it is common practice in reducing CCD images to compensate 
for a constant offset between magnitude scales of the images using one or
many reference stars \citep[e.g.][]{2016MNRAS.462.2506B,2017AJ....153...77C}, 
{\scshape VaST} uses a large number of matched stars to reconstruct 
{\it the scale offset as a function of magnitude.}
This allows {\scshape VaST} to calibrate magnitude scales of images obtained
with non-linear image detectors listed in Section~\ref{sec:intro}.
The only requirement is that the current and reference image have a
sufficient number of common stars ($\gtrsim 100$).
The dependency of the magnitudes in the instrumental scale of the reference
image, $m_{\rm ref}$, on the instrumental magnitudes of the current image,
$m_{\rm inst}$ (the calibration curve), is reconstructed using
all the stars matched between the two images and approximated by one of
 the three functions selected by the user:
\begin{enumerate}
\item linear function
\begin{equation}
\label{eq:linear} 
m_{\rm ref} = a_1 m_{\rm inst} + a_0
\end{equation}
\item parabola 
\begin{equation}
\label{eq:parabolic}
m_{\rm ref} = a_2 m_{\rm inst}^2 + a_1 m_{\rm inst} + a_0
\end{equation} 
\item ``photocurve'' function proposed by \cite{2005MNRAS.362..542B}:
\begin{equation}
\label{eq:photocurve}
m_{\rm ref} = a_0 \times \log_{10}\left({10^{a_1 \times (m_{\rm inst} - a_2)} + 1}\right) + a_3
\end{equation}
or the inverse function of it
\begin{equation}
\label{eq:invphotocurve}
m_{\rm ref} = {1 \over a_1} \times \log_{10}\left({10^{(m_{\rm inst} - a_3)\over a_0} - 1}\right) + a_2
\end{equation}
depending on which of the two functions provides a better fit to the data.
\end{enumerate}
Here $a_0$...$a_3$ are the free parameters of the fit. Fitting is
performed using the linear least-squares for the first two functions 
while the Levenberg-Marquardt algorithm is used 
to perform non-linear least-squares fitting of the photocurve.
The data points are weighted according to the inverse square of their
estimated photometric errors. An iterative clipping procedure is applied to
discard variable, poorly measured or misidentified objects.
An example magnitude scale calibration between two photographic images is
presented on the left panel of Figure~\ref{fig:magcalib}.

{\scshape VaST} can attempt to minimize the effect of the difference in
extinction across the image \citep[e.g.][]{2007MNRAS.375.1449I} 
by least-squares fitting a plane in the 3D dataset (X,Y,dm), where (X,Y) are
the image coordinates and dm is the residual difference in magnitudes
measured for a given object on the reference and the current image 
after applying the calibration curve.
The best-fit plane is then subtracted from the magnitudes measured on the
current image, thus correcting for the linear (in magnitude) term of the
extinction, regardless of image orientation. This correction is similar to
the coordinates correction performed after applying the initial
transformation derived from matched triangles (Section~\ref{sec:match}).
However, performing this correction may do more harm than good if the plane
cannot be fitted with sufficient accuracy. The correction may be
turned on or off by the user (\ref{sec:vastcommandlineopt}).
By default, the correction is applied to images
having $>10000$ detected sources. It is assumed that this large
number of sources will be sufficient to accurately fit the plane.
Note that this plane-fitting correction does not correct for
the extinction difference, but rather for the difference in extinction
difference between the reference and the current image.
Photometric calibration described in Section~\ref{sec:photometry} assumes one
zero-point for the whole field, resulting in offsets between
zero-points of lightcurves of objects visible in the upper and lower 
(with respect to the horizon) parts of the reference image.
At this stage {\scshape VaST} does not take into account the color term in
extinction correction (differential color extinction) as the colors of the
observed sources are, in general, unknown. 
After constructing lightcurves you may fit for the differential color 
extinction together with other systematic effects affecting multiple sources
in the field by applying the {\scshape SysRem} algorithm
as described in \ref{sec:examplesysrem}.

\subsection{Output lightcurves, statistics and the graphical interface}
\label{sec:outputgui}

The lightcurves of all the objects are saved in individual
\texttt{ASCII} files named \texttt{outNNNNN.dat} where \texttt{NNNNN} is the
source number.
Each line of the lightcurve file contains:
\begin{itemize}
\item The middle-of-exposure JD in TT or
UTC (Section~\ref{sec:fitsheader}).
\item Magnitude in the instrumental system corresponding to the reference
image.
\item Magnitude error estimated by {\scshape SExtractor}. The estimation
includes contributions from the background noise over the aperture area and
photon noise. The photon noise is estimated correctly only if 
the \texttt{GAIN} parameter is set correctly from the \texttt{FITS} header 
(Section~\ref{sec:fitsheader}) or by the user in the \texttt{default.sex} file.
\item Pixel coordinates of the object on the current image reported by
{\scshape SExtractor} (parameters \texttt{XWIN\_IMAGE} and \texttt{YWIN\_IMAGE}).
\item Aperture size in pixels used for this image.
\item Path to the image file from which the above measurements of the
object's parameters were obtained.
\end{itemize}

For each object {\scshape VaST} computes a number of variability indices
characterizing the scatter of magnitude measurements and/or smoothness of
the lightcurve \citep{2017MNRAS.464..274S}. 
The indices are constructed so that variable stars tend to have larger
values of the indices compared to the majority of non-variable stars, but the index values
(and their scatter) expected for non-variable stars are strong functions of magnitude, so the
simple cut in index values is usually not sufficient for efficient selection
of variable stars.
The index values for each object are stored in
\texttt{vast\_lightcurve\_statistics.log} its format being detailed in the
accompanying file \texttt{vast\_lightcurve\_statistics\_format.log}.
The index values may be utilized by the user as an input for automated selection of
candidate variables using external software implementing complex cuts in index
values or machine learning. 

\begin{figure}
 \centering
 \includegraphics[width=0.48\textwidth]{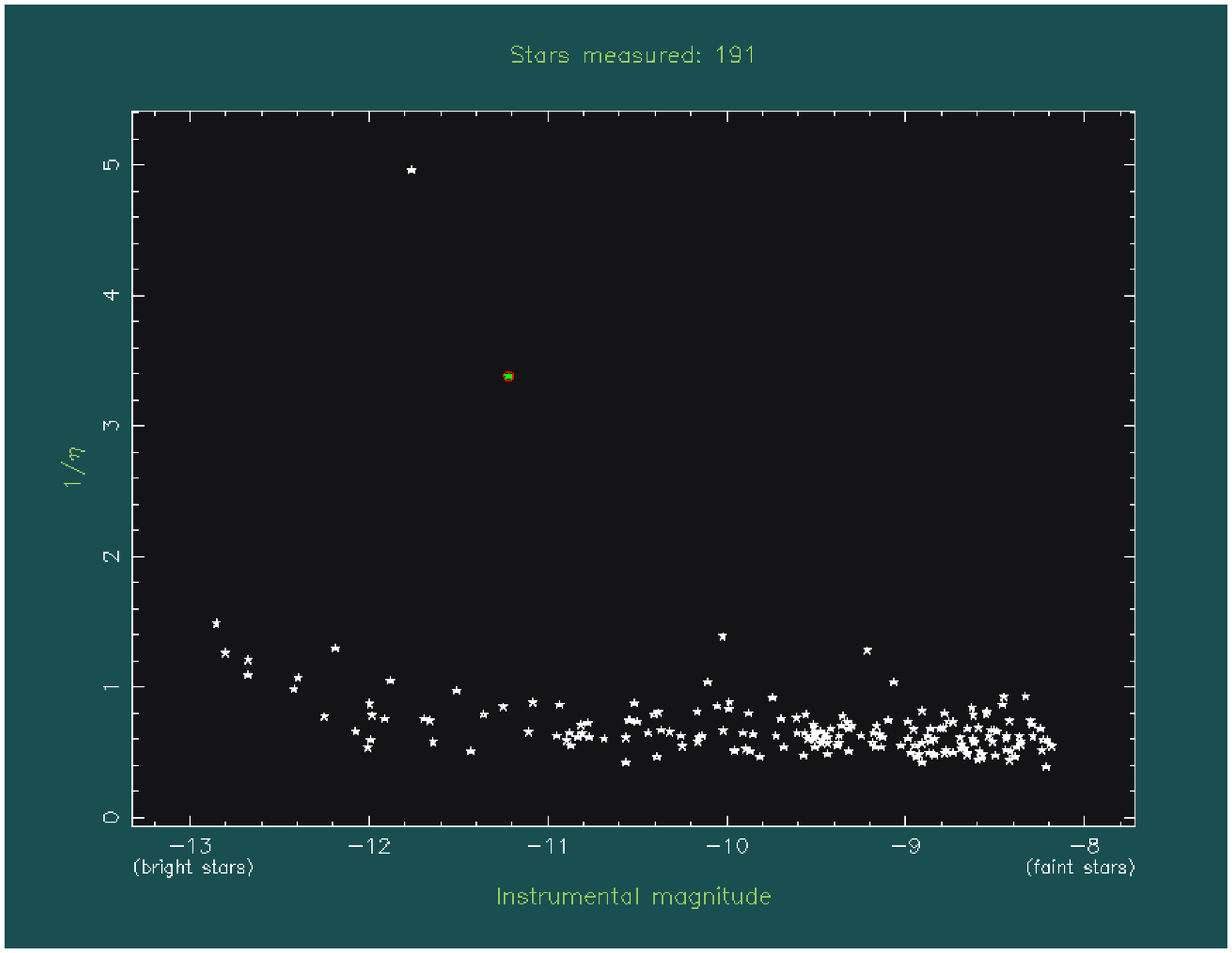}
 \includegraphics[width=0.48\textwidth]{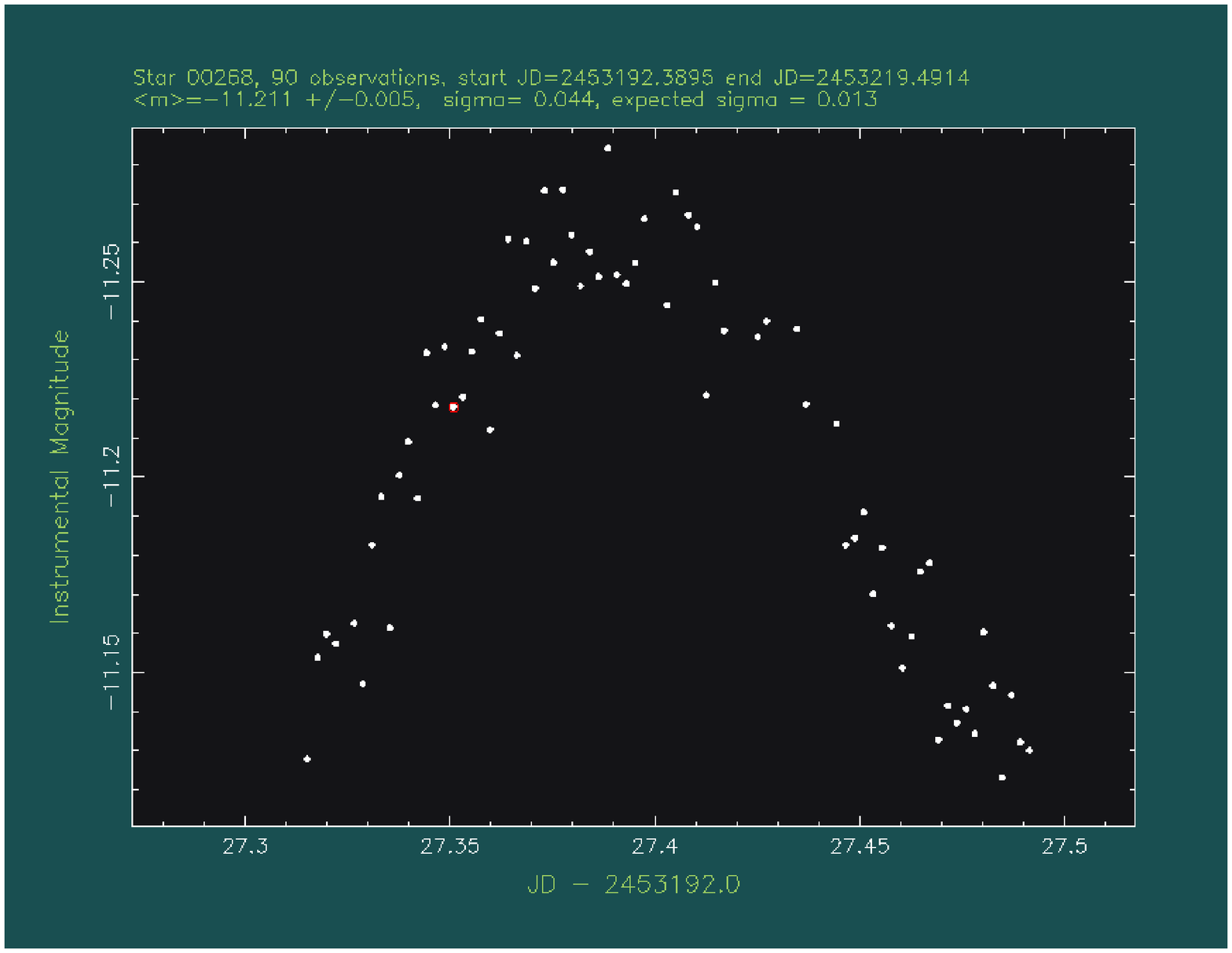}
 \includegraphics[width=0.48\textwidth]{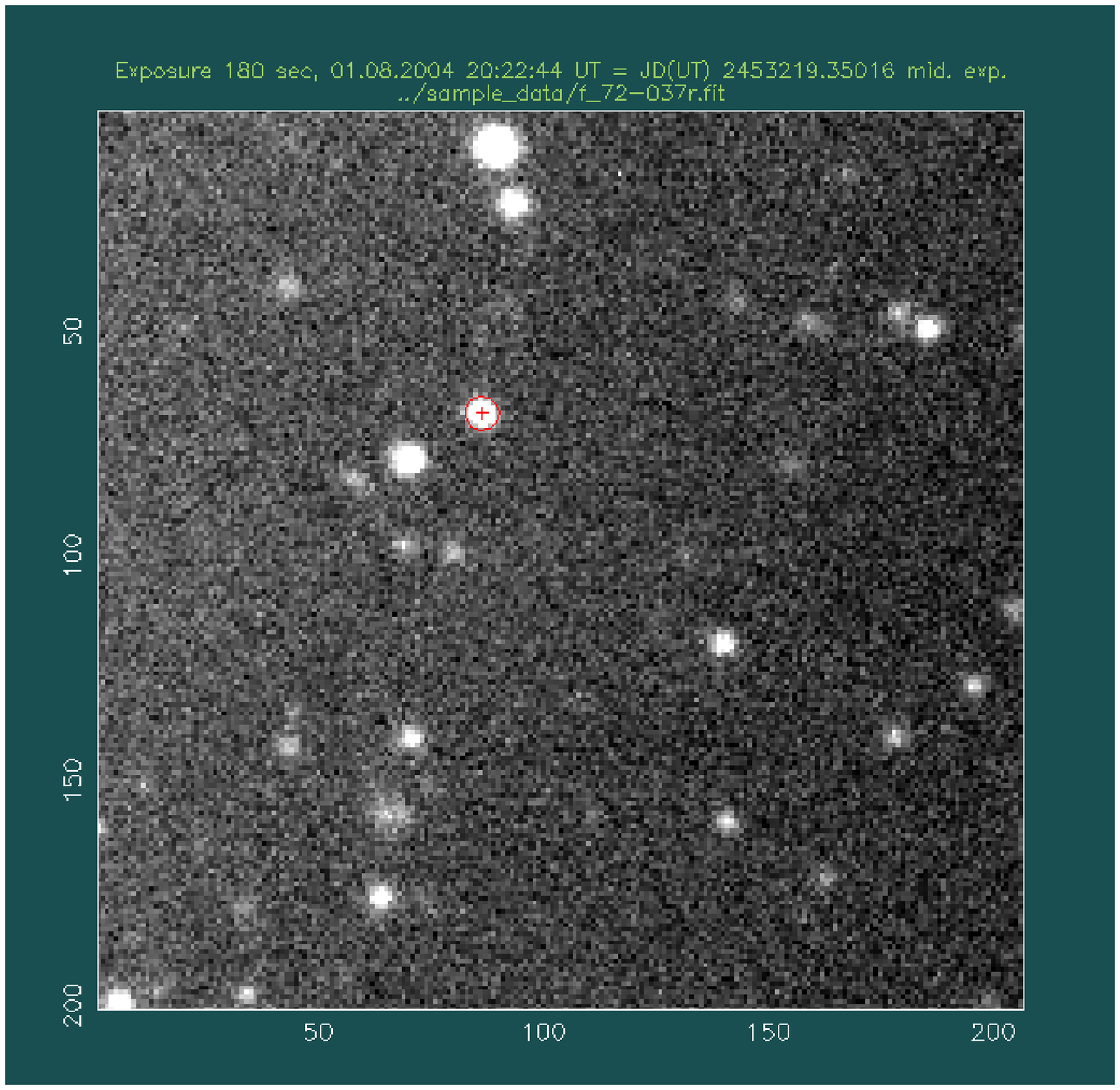}
 \caption{The {\scshape VaST} user interface (top to bottom): the main
window displaying a variability index--magnitude plot; the lightcurve
inspection window showing a star marked on the index--magnitude plot; 
\texttt{FITS} image viewer showing the image from which the lightcurve
point marked on the above plot was obtained.}
 \label{fig:vastgui}
\end{figure}

The simplest way of using {\scshape VaST} is to visualize the 
variability indices vs. magnitude plots using its GUI (Figure~\ref{fig:vastgui}).
By clicking on an object in these plots a user may visualize its lightcurve
and clicking on a point in the lightcurve plot -- display an image
corresponding to this point. This way it is possible to select objects
displaying high variability index values and make sure the apparent
variability is not caused by brightness measurement problems resulting from 
blending with nearby objects or CCD cosmetic defects (the problems that
readily appear on visual inspection of the object's images).
With one keystroke a user may launch the star identification script described in
Section~\ref{sec:astrometry} or send its lightcurve to the online period search 
tool\footnote{\url{http://scan.sai.msu.ru/lk/}}.

The {\scshape VaST} GUI is based on the {\scshape PGPLOT}
library\footnote{\url{http://www.astro.caltech.edu/~tjp/pgplot/}} 
which is well suited for displaying and editing data and image plots.
It is exceptionally easy to use for a developer. 
The downside is that the resulting interface is counterintuitive to a contemporary user
as it has no buttons, just the clickable plots. Whenever a user has a choice between multiple actions, 
instead of clicking a button to execute the desired action, a user has to press 
a key on a keyboard. For each GUI window, a list of possible keyboard keys is printed in
a terminal window. 

\subsection{Processing time}
\label{sec:processingtime}

{\scshape VaST} running time is mostly limited by the image processing time
with {\scshape SExtractor}. On a 2.20\,GHz quad core Intel Core~i7
laptop it takes about 2 minutes to process 784 images each containing about 250
stars (the dataset described in \ref{sec:softwaretest}) in the aperture photometry mode.
Processing the same images in the PSF-fitting mode takes about 50 minutes
using the same hardware.

\subsection{Astrometric calibration}
\label{sec:astrometry}

It is possible to construct lightcurves in the instrumental magnitude scale
and search for variability with {\scshape VaST} without finding the
transformation between the reference image and celestial coordinates.
If the field center is known to the user, detected variable stars can be
identified by visually comparing the displayed image with a star atlas like
{\scshape Aladin}\footnote{\url{http://aladin.u-strasbg.fr/}}
\citep{2000A&AS..143...33B}, see also \ref{sec:examplemanualmagcalib}.
However, this approach is practical only for narrow-field image sets
containing only few variable objects.

\begin{figure}
 \centering
 \includegraphics[width=0.48\textwidth]{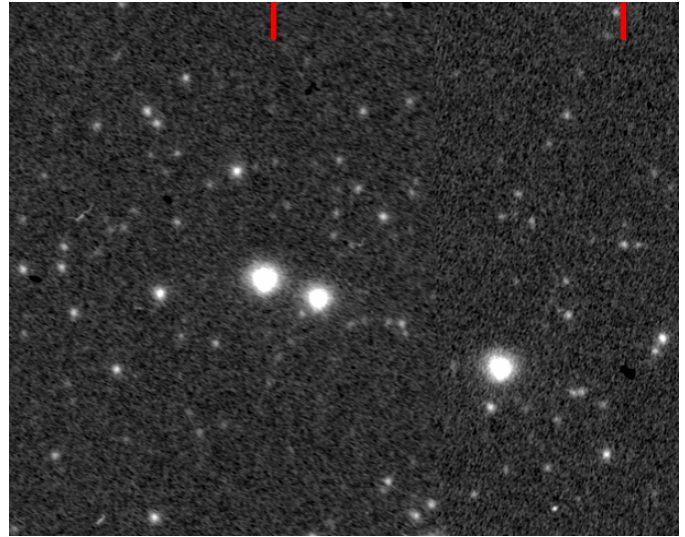}
 \caption{A section of a photographic plate digitized with a flatbed
 scanner. Red marks indicate the width of the stitch area
 between two passes of the scanning ruler (cf. left panel of Figure~\ref{fig:photoplateshifthacksaw}).}
 \label{fig:photoplatestitch}
\end{figure}

\begin{figure*}
 \centering
 \includegraphics[width=0.48\textwidth]{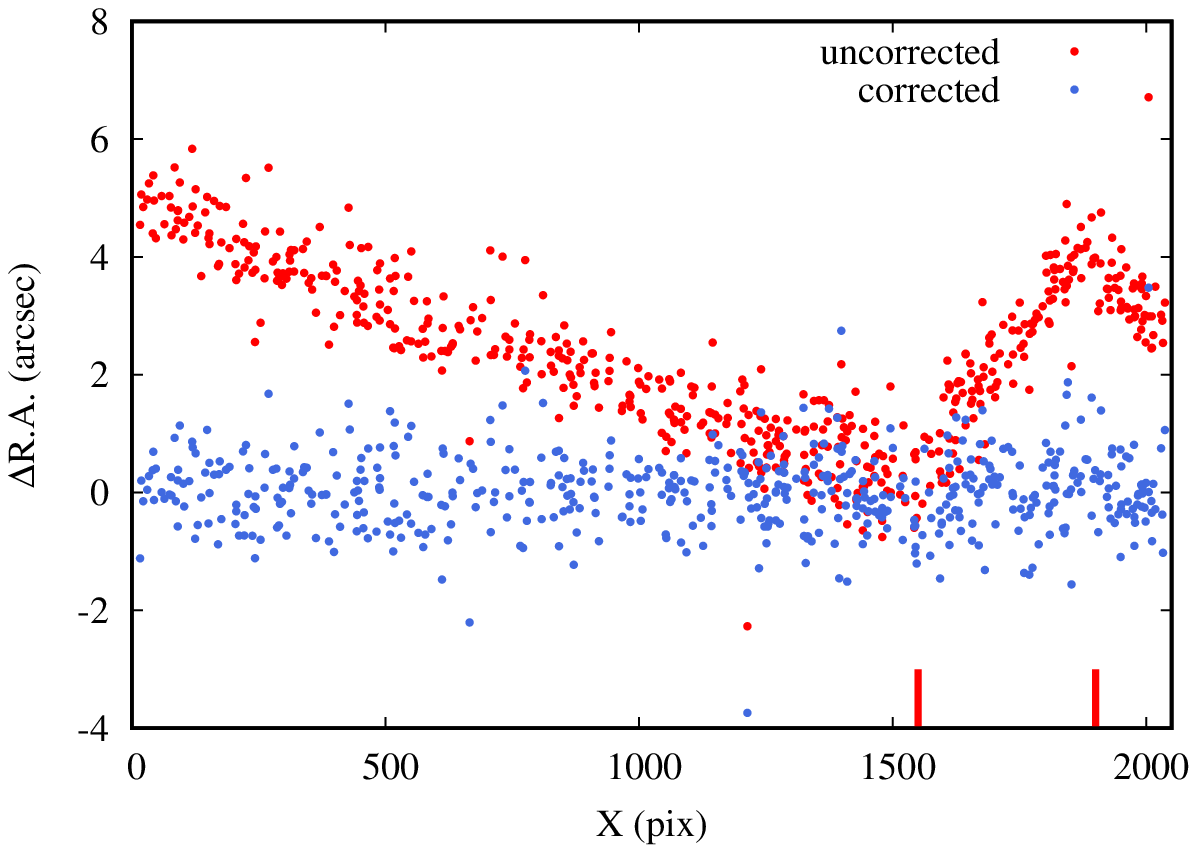}
 \includegraphics[width=0.48\textwidth]{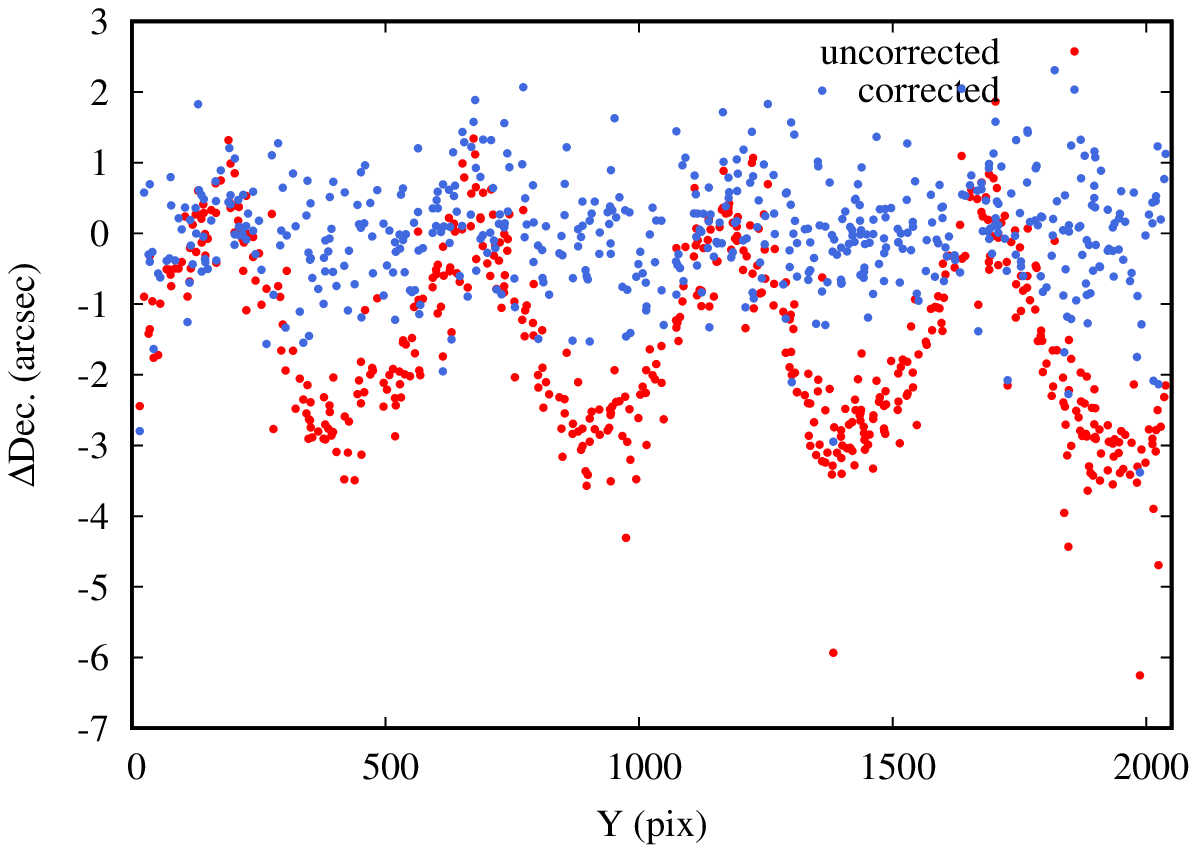}
 \caption{Differences between the cataloged and measured positions of stars
 on a digitized photographic plate. Distortion patterns commonly introduced 
 by flatbed scanners are illustrated: the stitch between two passes of 
 the scanning ruler (left; red bars indicate approximate boundaries of the stitch area, cf.~Figure~\ref{fig:photoplatestitch}) 
 and hacksaw distortions introduced by non-uniform motion of the ruler (right). 
 These distortions are mitigated by the local astrometric correction.}
 \label{fig:photoplateshifthacksaw}
\end{figure*}

For larger fields it is more practical to find a plate solution in an
automated way 
(see the usage example in \ref{sec:exampleautomagcalib}).
The script will use the {\scshape Astrometry.net} software to plate-solve 
the reference image and identify the star
with USNO-B1.0 \citep{2003AJ....125..984M} for positional reference and with
databases listing variable stars: GCVS\footnote{\url{http://www.sai.msu.su/gcvs/gcvs/}}
\citep{2009yCat....102025S}, VSX\footnote{\url{https://www.aavso.org/vsx/}} \citep{2006SASS...25...47W} and 
SIMBAD\footnote{\url{http://simbad.u-strasbg.fr/simbad/}}
\citep{2000A&AS..143....9W}.
The catalogs are accessed through {\scshape VizieR}\footnote{\url{http://vizier.u-strasbg.fr/}}
using {\scshape vizquery} or directly through a web interface using 
{\scshape cURL}.
The {\scshape Astrometry.net} software may be run by the script locally 
(if installed in the system) or on a remote server. In the latter case 
source extraction is done locally and only
the list of detected stars (not the full image) is uploaded to the server to save bandwidth.
The \texttt{FITS} header containing the WCS information created by {\scshape Astrometry.net}
is merged into the original \texttt{FITS} image and the approximate equatorial
coordinates of all sources are measured with the additional run of {\scshape SExtractor}.

In practice, the accuracy of the coordinates derived this way may not be
sufficient for unambiguous object identification for the following reasons.
While finding a blind plate solution {\scshape Astrometry.net} may fit for image
distortions and store them in the \texttt{FITS} header following the SIP
convention \citep{2005ASPC..347..491S}. This convention is not understood by
{\scshape SExtractor} which is using the PV convention to represent
geometric distortions \citep{2012SPIE.8451E..1MS}.
As a workaround, {\scshape VaST} employs {\scshape xy2sky} routine from 
{\scshape WCSTools} \citep{1997ASPC..125..249M,2014ASPC..485..231M} to convert {\scshape SExtractor}-derived
pixel coordinates to celestial coordinates rather than rely on 
{\scshape SExtractor} to do this conversion.

The following two problems are specific for large-format photographic plates digitized
with a flatbed scanner. If the scanner's image detector (scanning ruler) is 
smaller than the plate width, the detector has to make multiple passes
along the plate and the image strips resulting from each pass have to be
stitched together (Figure~\ref{fig:photoplatestitch}). Such a stitch results in a discontinuity in image to
celestial coordinates conversion (Figure~\ref{fig:photoplateshifthacksaw}, right panel) 
that cannot be adequately represented with a low-order polynomial description of distortions.
The other problem is the characteristic ``hacksaw'' distortions pattern
\citep{2007A&A...471.1077V}
resulting from non-uniform mechanical movement of the scanning ruler (see
Figure~\ref{fig:photoplateshifthacksaw}, left panel).
While the star matching algorithm proposed by \cite{2013MNRAS.433..935H}
is capable of dealing with shearing, we are not aware of an algorithm
that could accommodate a shift or a discontinuity in coordinates
transformation.

To minimize the negative effects of the above, {\scshape VaST} relies on the
assumptions that image distortions are similar for objects that are close to each
other and the distortions are sufficiently small to allow for correct
identification for the majority of objects.
After attempting to match all the detected objects with the UCAC4 catalog
\citep[][accessed from {\scshape VizieR}]{2013AJ....145...44Z} using the uncorrected coordinates, 
for each detected object {\scshape VaST} computes the mean difference between
the measured and the catalog positions of matched objects within 
a certain radius of the current object. This difference is
used as a {\it local astrometric correction}. 
A range of local correction radii is tested for each object and the one
resulting in the smallest scatter of the measured-to-catalog distances is used
to compute the final correction. 
The corrected positions of all the detected sources are used to match them with UCAC4 again
in an attempt to find new matches.
This procedure is repeated iteratively until the new iteration does not
result increased number of matched stars or the maximum number of iterations is reached.
The matching radius is set based on the image field of view.
Figure~\ref{fig:photoplateshifthacksaw} illustrates how the complex
distortions introduced by a flatbed scanner can be corrected by applying
the described procedure. The systematic effects are removed at the
expense of slightly increased random errors in astrometry.

\subsection{Photometric calibration}
\label{sec:photometry}

The magnitude scale can be calibrated manually using a nearby star. 
This can be done by interactively specifying a comparison star with a known magnitude as
described in \ref{sec:examplemanualmagcalib}.

If the reference image has a sufficiently large field of view to be blindly 
solved with {\scshape Astrometry.net}, the magnitude scale can be
calibrated using APASS \citep{2016yCat.2336....0H,2014CoSka..43..518H} magnitudes 
as described in \ref{sec:exampleautomagcalib}.
The following filters are supported: \texttt{B}, \texttt{V}, \texttt{R},
\texttt{I}, \texttt{r} and \texttt{i}.
The \texttt{R} and \texttt{I} magnitudes are not present in APASS, 
so they are computed from \texttt{r} and \texttt{i} magnitudes following \cite{2005AJ....130..873J}:
\begin{equation}
\label{eq:apassrbandconversion}
\texttt{R} = \texttt{V} - 1.09(\texttt{r}-\texttt{i}) - 0.22
\end{equation}
\begin{equation}
\texttt{I} = \texttt{R} - 1.00(\texttt{r}-\texttt{i}) + 0.21
\end{equation}
(the above relations already takes into account that \texttt{R} and
\texttt{I} magnitudes are defined in the Vega system while \texttt{r} and \texttt{i} magnitudes are always defined in
the AB system, \citealt{2005ARA&A..43..293B}).
%
%
The relation between the catalog magnitudes and the instrumental magnitudes
is approximated by one of the relations (\ref{eq:linear})--(\ref{eq:invphotocurve})
selected by the user.
An example relation between the catalog and measured instrumental magnitudes is
presented on the right panel of Figure~\ref{fig:magcalib}.
{\scshape VaST} assumes the magnitude calibration curve is equally applicable to all
objects across the image ignoring the effects of differential extinction
(which are important for wide-field images) and possible flat-fielding
imperfections.

\subsection{Flag images}
\label{sec:flagimg}

Sometimes input \texttt{FITS} images contain large regions not covered by data: 
overscan columns, gaps between chips of a mosaic CCD and an area around 
the originally rotated image after it was resampled to the North-up/East-left orientation
(Figure~\ref{fig:flagimage}).
Sources detected on the boundary between these areas and areas covered by
data are not flagged by internal {\scshape SExtractor} flags, as the program
has no way of knowing that these are the physical edges of an image.
{\scshape VaST} tries to identify large clusters of zero-value
pixels (isolated zero-value pixels may be common if the image is background-subtracted)
and produces a flag image that is masking these clusters and a few pixels
surrounding them. The generated flag image is supplied to {\scshape SExtractor}
and used to filter-out sources close to the image edges.
An example flag image is presented in the right panel of Figure~\ref{fig:flagimage}.
Note that the flag image is used only to flag the spurious detections.
It does not affect background level calculations performed by 
{\scshape SExtractor}.

\begin{figure}
 \centering
 \includegraphics[width=0.22\textwidth]{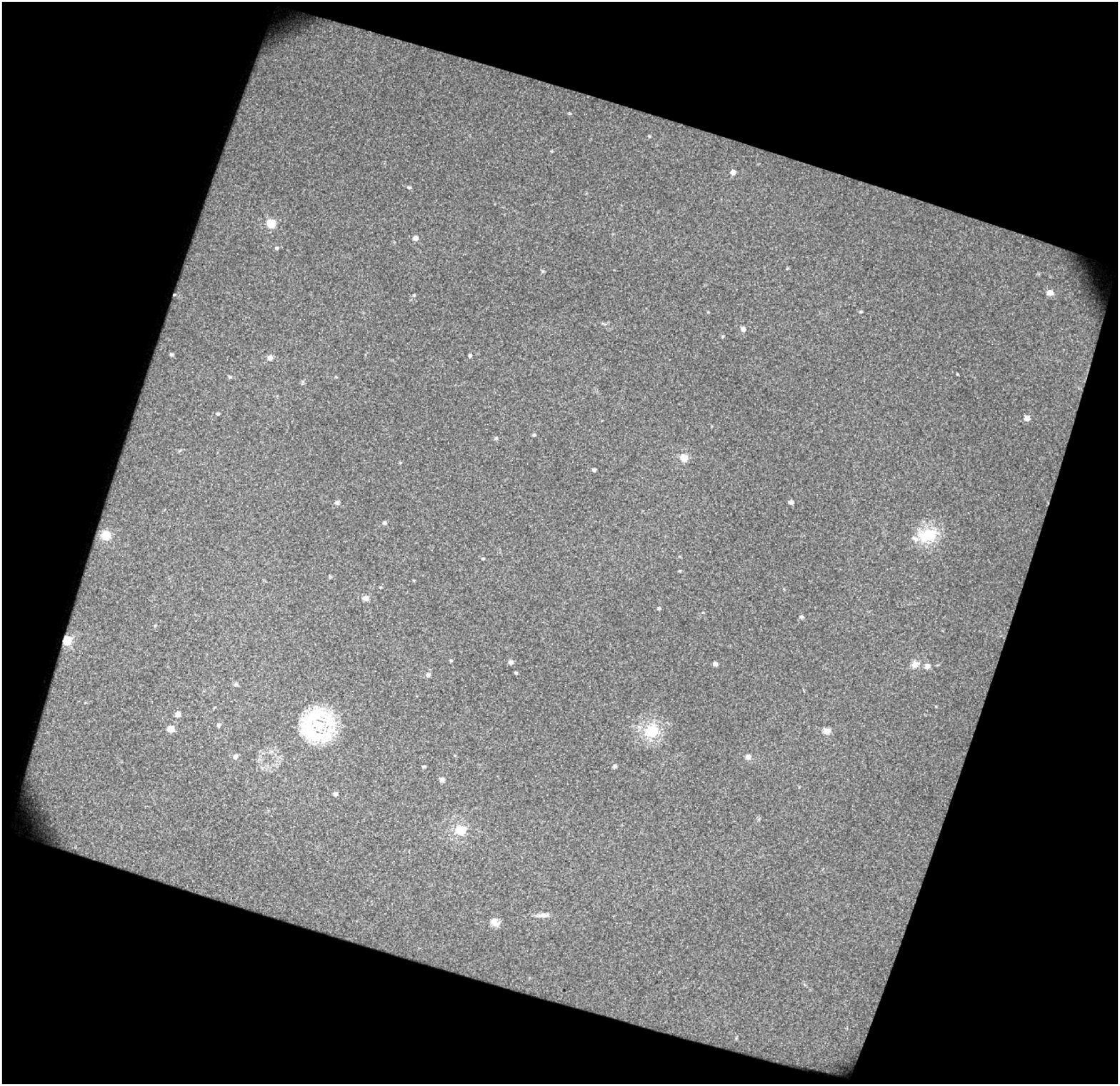}
 \includegraphics[width=0.22\textwidth]{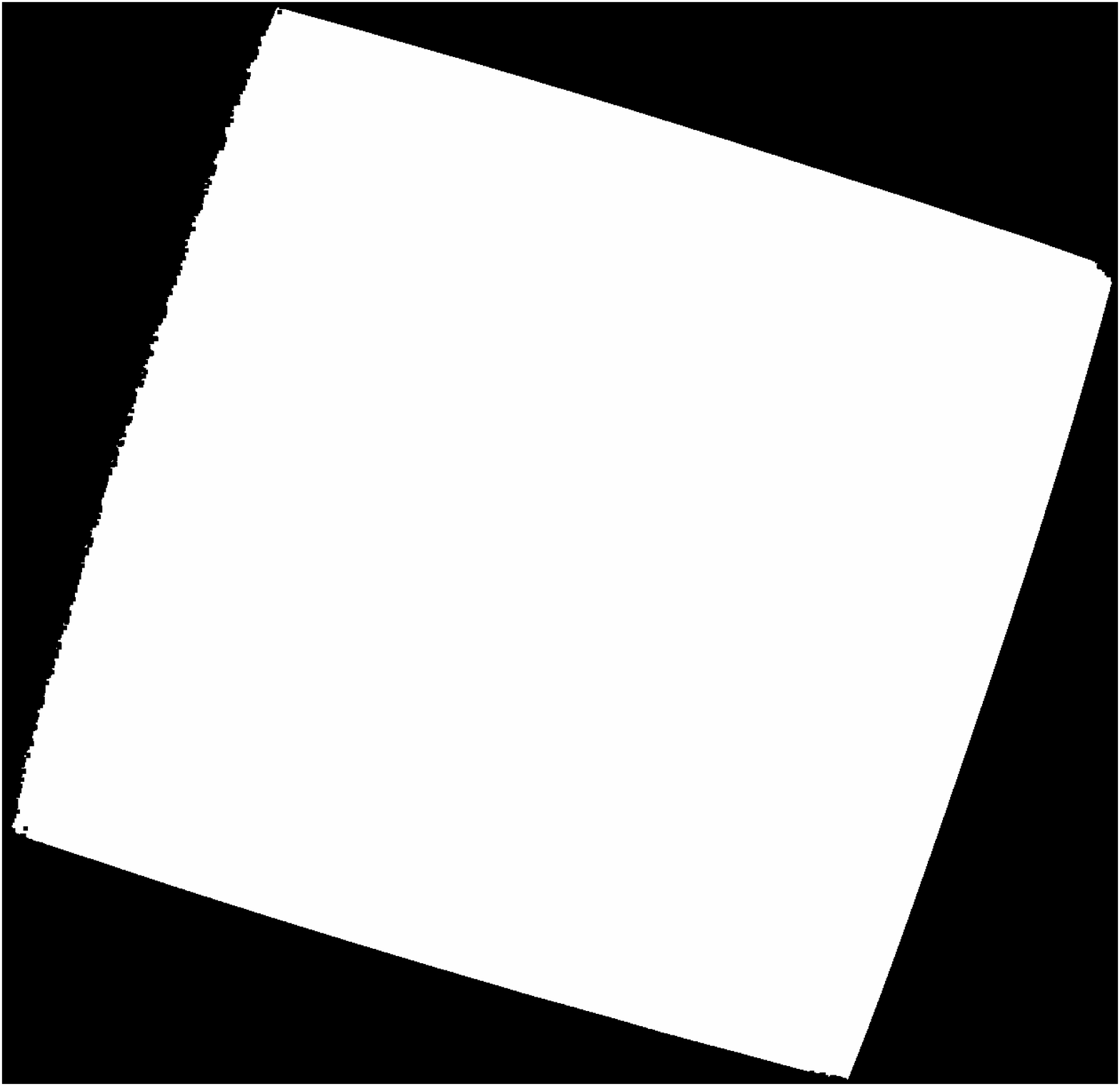}
 \caption{An individual exposure extracted from Swift/UVOT Level~2 image (left) and the corresponding flag image
 created by {\scshape VaST} (right).}
 \label{fig:flagimage}
\end{figure}

\section{Searching for transients with {\scshape VaST}}
\label{sec:transientssec}

\begin{figure*}
 \centering
 \includegraphics[width=0.9\textwidth]{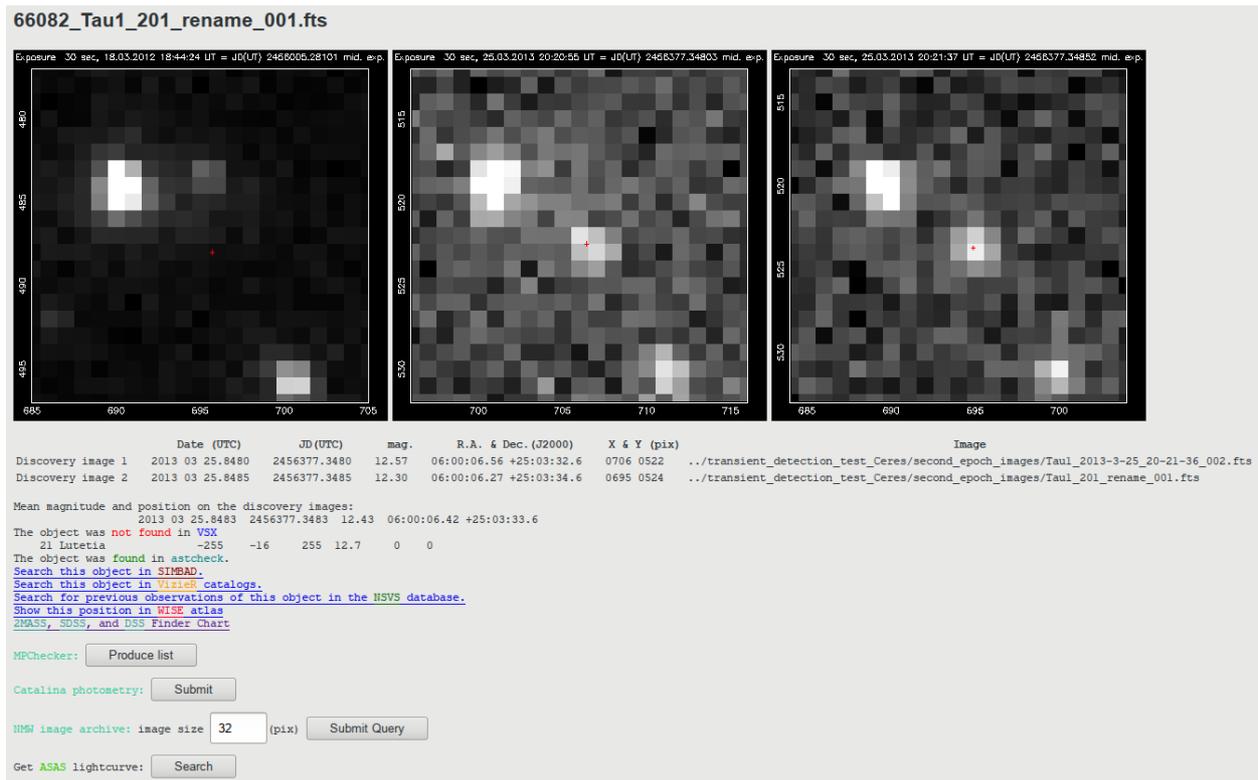}
 \caption{Transient candidates report page viewed in a web browser. This
candidate is identified as the asteroid (21)~Lutetia.}
 \label{fig:lutetia}
\end{figure*}

Transient sources like supernovae, novae and dwarf novae 
are often found with the image subtraction technique \citep{1998ApJ...503..325A,2016MNRAS.457..542B}. 
Unlike a small-amplitude variable star, a new object appearing well above 
the detection limit results in an obvious change in an image of a star field. 
A list of software implementing this technique may be found in Section~\ref{sec:intro}.
Image subtraction has the advantage that it can naturally handle a new object appearing 
on top of a previously visible one, like a supernova appearing on top of
a galaxy image. With the traditional source detection on sky images (rather
than on difference images) to learn that something has changed the supernova has 
to be either successfully deblended from the host galaxy or it should be
sufficiently bright to cause a detectable change in the measured brightness
of the galaxy$+$supernova compared to the measured brightness of the galaxy alone. 
However, 
the image subtraction may be difficult to implement 
if the PSF and its variations across the images are difficult to reconstruct.
In such cases transient detection based on comparison of the lists of
sources detected on first- and second-epoch images may still be preferable.

The transient detection was implemented in {\scshape VaST}
for processing the New Milky Way (NMW\footnote{\url{http://scan.sai.msu.ru/nmw/}}) 
nova patrol \citep{2014ASPC..490..395S} images. 
Unlike the main lightcurve-based variability search mode that is fairly generic,
{\scshape VaST}'s transient detection mode is applicable only to wide-field
relatively shallow images (as it relies on Tycho-2 catalog for magnitude
calibration; \citealt{2000A&A...355L..27H}) and tied to the specific observing strategy.
For each transient survey field, exactly four images are required as the
input: tow first-epoch (reference) images and two second-epoch images. The two
images at each epoch should be obtained with a sufficiently large shift 
(20 pixels or more) to suppress spurious detections due to image artifacts.
Two first-epoch images are needed to reduce the probability of an object
visible on the reference image not being detected by {\scshape SExtractor}
due to blending or an image artifact. If this object is detected on the
second-epoch images - it would be incorrectly identified as new.

The candidate transients are selected as objects that were not detected at
any of the reference images, or where at least 1 mag. fainter compared 
to the second-epoch images.
The second criterion is needed to identify flaring objects and new objects
that are blended with previously-visible ones.

{\scshape VaST} generates an \texttt{HTML} report containing a list of candidate transients
and opens it in a web browser.
%
For each candidate transient the \texttt{HTML} report displays the following information
(Figure~\ref{fig:lutetia}):
\begin{itemize}
\item Cutouts from the first and second epoch images centered on the transient candidate.
\item Photometry and astrometry of the candidate.
\item Results of VSX search for known variable stars around the transient's
position. A local copy of the catalog is used.
\item Results of {\scshape astcheck}\footnote{\url{http://www.projectpluto.com/pluto/devel/astcheck.htm}} 
search for known asteroids around the transient's position; {\scshape astcheck}
relies on a local copy of asteroid orbit data.
\item Links to search the transient's position in {\scshape SIMBAD} and {\scshape VizieR} databases as well as 
NSVS, ASAS \citep{1997AcA....47..467P,2002AcA....52..397P} and CSS \citep{2009ApJ...696..870D}
photometry archives. A link to the WISE \citep{2010AJ....140.1868W} image atlas is also
provided. It is especially useful for distinguishing unknown red variables (bright in
the infrared light) from nova/cataclysmic variable candidates. 
The object's position may also be checked in MPChecker\footnote{\url{http://www.minorplanetcenter.net/}}
(that has the latest information on asteroids and comets) and the NMW image
archive. 
\end{itemize}
%
%
\ref{sec:transients} provides an example of running {\scshape VaST} in the
transient detection mode.

\section{Remarks on development process}
\label{sec:remarks}

Some of the best practices for scientific computing were reviewed by
\cite{2012arXiv1210.0530W}. Here we highlight a few procedures 
that were found especially useful for {\scshape VaST} development.

The core functionality of {\scshape VaST} is implemented in {\scshape C}.
{\scshape Valgrind}\footnote{\url{http://valgrind.org/}} \citep{Net:valgrind2007} tools are used for 
profiling \citep{Net:workload-characterization2006} and
detecting memory errors and leaks \citep{Sew:memcheck2005}.
Parallel processing is implemented using OpenMP.

{\it Defensive programming} style is adopted whenever possible.
{\scshape VaST} continues execution after failing to process an individual
image. With this we are trying to avoid the situation when hours of computation 
are lost because of a bad image mixed into a generally good input dataset
\citep[e.g.][]{2001ASPC..238..306R}.
An attempt is made to print a meaningful error message in cases when
the execution of the program cannot continue: most such situations are caused
by incorrect input on the command-line or from image/lightcurve files and
can be corrected by the user.

{\it Automated testing} was not implemented from the start of the project,
but it quickly appeared that adding new functionality to the code broke
some of the rarely-used functionality added earlier. The solution was found
in system testing implemented in a form of a {\scshape BASH} script.
The script runs {\scshape VaST} in a non-interactive mode on various sets of
test data and checks if the output is consistent with the expectations.
After {\scshape VaST} processed a set of test images, the script checks if
all middle-of-exposure JDs were computed correctly, 
if all the images were successfully matched to the reference image, if the
reference image was plate-solved and if the object having the highest
lightcurve scatter can be identified with a known variable star or if 
a known flaring or moving object is among the list of detected transients.
While the tests check high-level functionality of the software, 
once a problem is discovered it is often easy to pin-point a recently
modified piece of code that caused it.
Bug reports from {\scshape VaST} users provide a steady source of non-trivial test cases.

{\scshape VaST} is developed and tested under {\scshape Gentoo Linux} that,
in the authors' view, provides a developer-friendly environment. 
A set of portability tests is performed prior to release. 
The tests ensure that {\scshape VaST} compiles and runs well on
the latest {\scshape Ubuntu} and
{\scshape Scientific Linux}, as well as the old 
{\scshape Scientific Linux~5.6} 
recommended\footnote{\url{https://boinc.berkeley.edu/trac/wiki/VmCompatibility}} 
for building portable applications by the {\scshape BOINC}
project \citep{2012AREPS..40...69K}.
The testing is also performed on the latest stable release of {\scshape FreeBSD}.
All these operating systems are run as virtual machines through {\scshape VirtualBox}.
A number of bizarre differences in behavior between versions of 
{\scshape gcc}, {\scshape make}, {\scshape BASH}, {\scshape awk} and {\scshape sort}
commands supplied with different {\scshape Linux} distributions 
(not to mention differences between their version supplied with 
{\scshape Linux} and {\scshape FreeBSD}) were encountered, 
confirming the need for the portability testing\footnote{For example, 
{\scshape mawk} which is the default {\scshape awk} in {\scshape Ubuntu} will not
understand \texttt{"\%lf"} in the \texttt{printf} formatting string
(accepting only \texttt{"\%f"}):\\
\texttt{echo 1.23 | awk \char"0D\{printf "\%lf\textbackslash n",\$1\}\char"0D}\\
while it is perfectly fine for {\scshape gawk} which is the
default in most {\scshape Linux} distributions as well as {\scshape BSD awk}.
This kind of differences are hard to anticipate, so they need to be tested
for.}.
We emphasize that the operating system versions mentioned above are 
not the ones strictly required for running {\scshape VaST}, rather they
should be representative examples of the current UNIX-like systems diversity.
The goal of our portability testing is to have a reasonable expectation 
that {\scshape VaST} will compile and run on {\it any} contemporary 
{\scshape Linux} or {\scshape FreeBSD} system.

To make {\scshape VaST} installation as easy as running the command
\texttt{make}, the package includes copies of
some non-standard libraries it relies on that are automatically built from 
source code before compiling {\scshape VaST} source files.
{\scshape VaST} also comes with its own copy of 
{\scshape SExtractor} however it is configured with 
\texttt{--disable-model-fitting} to avoid dependency on the 
{\scshape ATLAS} library and therefore cannot perform PSF-fitting.
To enable PSF-fitting photometry one has to install 
{\scshape ATLAS}, {\scshape PSFEx} and {\scshape SExtractor} system-wide.
{\scshape VaST} will use the system installation of {\scshape SExtractor}
if it finds one.

The main obstacle in {\scshape VaST} adaptation by the users appears to be
the complexity of its command-line interface combined with the lack of clear
documentation. While software with overly complex user interfaces and
non-trivial installation procedures (think {\scshape AIPS}, {\scshape ESO-MIDAS}, 
{\scshape IRAF}) are often tolerated in astronomy for historical reasons,
many variable star observers expect software in this field to have an
intuitively-understandable GUI. The problem is not confined to the interface
of the program, but also includes the operating system interface.
Familiarity of potential users with 
{\scshape POSIX}-like 
command-line interface is increasing thanks to the rise in popularity of 
{\scshape Mac~OS}. {\scshape VaST} documentation should be improved, in
particular video tutorials seem to be a good way to introduce the software
to potential users. We are also considering a radical re-design of the user interface 
to make it fully web-based.


\section{{\scshape VaST} applications}
\label{sec:applications}

The development of {\scshape VaST} was mostly driven by the author's data processing
needs. These evolved from a piggyback variability search in targeted
CCD observations of known variable objects \citep{2005IBVS.5654....1A,2011PZ.....31....1S}
to variability search using digitized photographic plates
\citep{2008AcA....58..279K,2014ARep...58..319S}, 
Swift/UVOT data analysis \citep{2009PZP.....9....9S,2012MNRAS.425.1357G},
visualization of CoRoT lightcurves \citep{2010CoAst.161...55S},
photometry of individual active galactic nuclei \citep{2011A&A...532A.150S,2014A&A...565A..26S}
and wide-field search for bright optical transients \citep{2014ASPC..490..395S}
that resulted in the discovery of Nova Sagittarii 2012~\#1 
\citep[=V5589~Sgr;][]{2012CBET.3089....1K,2012ATel.4061....1S}.
{\scshape VaST} was applied for variability search in CCD data by others including
\cite{2013A&A...560A..22M,2014PZP....14...12L,2015PZP....15....3K}.
\cite{2017arXiv171007290P} used {\scshape VaST} to compute multiple
variability indexes needed to test a machine learning-based approach to
variable star detection.
From the list of {\scshape VaST}-related publications maintained at the code's 
homepage\footnote{\url{http://scan.sai.msu.ru/vast/\#public}} it is
estimated that {\scshape VaST} contributed to the discovery of 
at least 1800 variable stars.

\section{Summary}
\label{sec:summ}

{\scshape VaST} takes a series of sky images as an input and produces
lightcurves for all imaged objects. A set of variability indices is computed
that characterize scatter and smoothness of the lightcurves and allow the
user to distinguish variable objects from non-variable ones. The user
interface aids in visual inspection of candidate variables by visualizing 
variability index--magnitude plots, lightcurves and images associated with 
each lightcurve point. Thanks to parallel processing and smart memory
management, the code is able to process thousands of images while
running on modest hardware.
{\scshape VaST} is free software, distributed under the terms of the GNU General Public
License (GPL)\footnote{\url{https://www.gnu.org/licenses/gpl.html}}.
We hope that the description of the code provided here will be useful both to
{\scshape VaST} users and those who aim to develop the next generation 
variability search software.

\section*{Acknowledgments}
We thank
the anonymous referees, 
Dr.~Alceste Bonanos,
George Kakaletris, 
Maria Mogilen,
Dean Roberts,
Dr.~Nikolay Samus,
Olga Sokolovskaya,
Dr.~Alexandra Zubareva
and
Dmitry Litvinov
for critically reading the manuscript.
Thanks to Mark Vinogradov for the suggestion on improving the text.
We thank Dmitry Nasonov who designed the {\scshape VaST} homepage,
Sergei Nazarov (Moscow) who coined the name for the code and
Sergei Nazarov (CrAO) who created great 
instructions\footnote{\url{http://astrotourist.info/poisk-peremennykh-zvezd}} 
on variability search (in Russian), 
Stanislav Korotkiy for starting the NMW nova patrol project,
Dr.~Alexei Alakoz, Dr.~Panagiotis Gavras and
Dr.~Jean-Baptiste Marquette
for testing the code on {\scshape Mac~OS}.
KS thanks Dr.~Sergei Antipin and Dr.~Vladimir Amirkhanyan 
for illuminating discussions of photometric data analysis, 
Daria Kolesnikova and Andrey Samokhvalov for many suggestions 
that helped to improve the output and interface of the program,
Dr.~Richard White for the illuminating discussion of flag images 
and weight maps handling by {\scshape SExtractor}.
This work is partly based on observations made with the 2.3\,m~Aristarchos telescope, Helmos Observatory, Greece,
which is operated by the Institute for Astronomy, Astrophysics, Space Applications and Remote
Sensing of the National Observatory of Athens, Greece.
This research has made use of the {\scshape SIMBAD} database,
operated at CDS, Strasbourg, France.
This research has made use of the {\scshape VizieR} 
catalog 
access tool, CDS,
Strasbourg, France. The original description of the {\scshape VizieR} service is
presented by \cite{2000A&AS..143...23O}.
This research has made use of the International Variable Star Index (VSX)
database, operated at AAVSO, Cambridge, Massachusetts, USA.
This research has made use of NASA's Astrophysics Data System.

\appendix

\section{Use cases}
\label{sec:usecases}

This section provides installation instructions and practical examples of 
how common variability-search problems can be solved with {\scshape VaST}.

\subsection{Compiling {\scshape VaST}}
\label{sec:compilingvast}

To run {\scshape VaST} you will need a computer running 
{\scshape Linux}, {\scshape FreeBSD} or {\scshape Mac~OS~X} operating
system. If you have a different system -- run one of the supported
systems in a virtual machine. Make sure {\scshape gcc} compiler with {\scshape C++}
and {\scshape Fortran} support as well as header files needed to compile 
{\scshape X.Org} GUI applications are installed. 

In a terminal window, download and unpack the archive containing {\scshape VaST} source
code, change to the unpacked directory and compile the code:
\begin{lstlisting}[language=novastmarkup]
wget http://scan.sai.msu.ru/vast/vast-1.0rc78.tar.bz2
tar -xvf vast-1.0rc78.tar.bz2
cd vast-1.0rc78
make
\end{lstlisting}
The script may complain about missing external programs or header files.
Install any missing components in your system and run \texttt{make} again. 
At this point you should have a working version of {\scshape VaST} 
ready to perform aperture photometry, as descried in the following examples.
All processing is done from within the {\scshape VaST} directory (\texttt{vast-1.0rc78}
in the above example) -- no system-wide installation is needed.


\subsection{Variability search in a series of CCD images}
\label{sec:simpleccdsearch}

Suppose we have a series of dark-subtracted and flat-field-corrected images
of a particular star field in the directory \texttt{../sample\_data}
To construct lightcurves and perform variability search in these images,
form the {\scshape VaST} directory run
\begin{lstlisting}
./vast ../sample_data/*
\end{lstlisting}
Here the input images are specified as command-line arguments and 
the \texttt{*} sign indicates all files in the directory.
The files that are not \texttt{FITS} images will be automatically ignored. 
After performing the processing steps described in Sections~\ref{sec:fitsheader}--\ref{sec:calibinstmag}, 
the program will
open an interactive window displaying a variability index--magnitude plot 
and allowing to inspect individual lightcurves and images as described 
in Section~\ref{sec:outputgui}.

\subsection{Setting a reference image of your choice}
\label{sec:refimage}

{\scshape VaST} is using the first image specified on the command-line as
the reference image. This may be a bad choice if the first image in a series
is of poor quality. To specify a different image, just put it first on the
command-line. In the example above, one may set
\texttt{f\_72-058r.fit} as the reference image
\begin{lstlisting}
./vast ../sample_data/f_72-058r.fit ../sample_data/*
\end{lstlisting}
Note that in this example \texttt{f\_72-058r.fit} appears on the list of
input images twice: the first time it is specified explicitly and the second
time -- when it satisfies the condition specified with the wildcard
character \texttt{*}. {\scshape VaST} will recognize this is again the
reference image and will not try to measure it twice.

\subsection{Restarting work after a break}
\label{sec:restartwork}

The main interactive window may be closed by right-clicking or pressing 'X'
on the keyboard twice. It may be re-opened with
\begin{lstlisting}
./find_candidates -a
\end{lstlisting}
If one runs \texttt{./find\_candidates} without a command-line argument, it
will re-compute the variability indices instead of re-displaying the old
computations. This is useful if the lightcurves
were altered, e.g. after calibrating the magnitude scale
(\ref{sec:examplemanualmagcalib}, \ref{sec:exampleautomagcalib})
or applying {\scshape SysRem} algorithm (\ref{sec:examplesysrem}).

All the lightcurves and log files produced after measuring a set of images
may be saved to a directory \texttt{MY\_FIELD\_NAME}
\begin{lstlisting}
util/save.sh MY_FIELD_NAME
\end{lstlisting}
and loaded back to the {\scshape VaST} working directory
\begin{lstlisting}
util/load.sh MY_FIELD_NAME
\end{lstlisting}

\subsection{View a lightcurve of an individual star}

The user may view a lightcurve of an object by selecting it on the
reference image:
\begin{lstlisting}
./select_star_on_reference_image
\end{lstlisting}
or if the object's number (12345) is known from the previous {\scshape VaST} run,
the lightcurve can be plotted with
\begin{lstlisting}
./lc out12345.dat
\end{lstlisting}
You may zoom in to a section of the lightcurve by pressing \texttt{'Z'} on 
the keyboard (press \texttt{'Z'} twice to zoom out). 
You may send the entire lightcurve to the online period search tool by
pressing \texttt{'L'}. If the lightcurve is generated by {\scshape VaST} and
not imported from external software, you may click on each lightcurve point 
to view the image corresponding to this point. You may try to identify the
object with a known variable by pressing \texttt{'U'} (see Section~\ref{sec:astrometry}).
As with the other {\scshape VaST} GUI applications, look at the terminal to
see the list of possible keyboard commands.

\subsection{Manual single-star magnitude calibration}
\label{sec:examplemanualmagcalib}

After generating lightcurves from a set of images 
(\ref{sec:simpleccdsearch}, \ref{sec:psffitting} or
\ref{sec:photoplateexample}) run
\begin{lstlisting}
util/magnitude_calibration.sh
\end{lstlisting}
The script will display an image marking the stars that pass all the
quality cuts. Identify a reference star of your choice by clicking on it and
enter its magnitude in the terminal. The APASS catalog visualized through {\scshape Aladin}
is a good place to find reference stars 
(Section~\ref{sec:photometry}).
An example of such single-star zero-point calibration is discussed in \ref{sec:softwaretest}. 

\subsection{Automated magnitude scale calibration}
\label{sec:exampleautomagcalib}

If the field of view is large enough to be automatically plate-solved,
the magnitude calibration can be performed by automatically matching the
detected stars to the APASS catalog:
\begin{lstlisting}
util/magnitude_calibration.sh V
\end{lstlisting}
where the command-line argument specifies the observing band (see
Section~\ref{sec:photometry}).
An interactive plot displaying the APASS magnitude as a function of 
the instrumental magnitude (Figure~\ref{fig:magcalib}) will be displayed. 
A fitting function may be chosen among the ones described in
Section~\ref{sec:calibinstmag} by pressing 'P'. 
Weighting may be turned on or of by pressing 'W'. Outlier points may be
interactively removed by pressing 'C' and drawing a rectangle with a mouse
around the outliers. After a satisfactory fit is found, right-click to apply 
the calibration.

\subsection{Fine-tuning source extraction parameters}
\label{sec:finetuningsourceextraction}

To check how well star detection is performed on an image:
\begin{lstlisting}
./sextract_single_image ../sample_data/f_72-001r.fit
\end{lstlisting}
where \texttt{../sample\_data/f\_72-001r.fit} is the path to the image file.
This will display the image marking all the detected objects with green
circles. Normally, one would like to have all objects clearly visible on the image
to be detected while a minimal number of obvious artifacts like hot pixels 
or noise peaks to be mistaken for real astronomical sources. The default source
extraction parameters work well for most CCD images while the images
acquired with DSLR cameras (typically equipped with CMOS detectors) and
digitized photographic images typically require fine-tuning of the extraction
parameters.

The extraction parameters are set in the \texttt{default.sex} file.
The most relevant are 
\texttt{DETECT\_MINAREA} and 
\texttt{DETECT\_THRESH}.
A detailed description of the parameters may be found in 
{\scshape SExtractor} 
documentation\footnote{\begin{sloppypar}\url{https://www.astromatic.net/pubsvn/software/sextractor/trunk/doc/sextractor.pdf}\end{sloppypar}}.
The {\scshape VaST} directory includes a few example files named \texttt{default.sex*}.

A click on a detected source will print its derived properties (including
pixel position and instrumental magnitude) in the terminal.
Inspection of object's properties is useful to find out why a particular 
known object of interest did not pass the selection criteria and its lightcurve was not generated.
It is useful to check {\scshape SExtractor} flags typically assigned to the detected objects.
{\it By default, {\scshape VaST} only accepts sources that have flag values
0 or 1.} If the star field is very crowded or the instrument's PSF has a
funny shape, the majority of objects may be marked with flags 2 or 3.
In that case you may either try to find more optimal extraction parameters
by changing the detection threshold as described above as well as
deblending parameters \texttt{DEBLEND\_NTHRESH} and \texttt{DEBLEND\_MINCONT}
or run {\scshape VaST} with \texttt{-x3} command-line argument in order not
to filter out blended stars. 

\subsection{Selecting best aperture size for each source}
\label{sec:varapsize}

{\scshape VaST} may use {\scshape SExtractor} to measure brightness of each
source in multiple circular apertures. For each source all apertures have the 
same center and their sizes are 10\,per~cent smaller, equal, 10. 20 and 30\,per~cent
larger than the size of the reference aperture selected automatically for each image 
based on seeing (as described in Sec.~\ref{sec:sextraction}) or set to a fixed size 
by the user for the whole series of images (\ref{sec:vastcommandlineopt}). 
After processing all the images, {\scshape VaST} may select for each source the aperture that
resulted in the smallest lightcurve scatter (quantified by MAD). 
The following command will activate the best aperture selection:

\begin{lstlisting}
./vast --selectbestaperture ../sample_data/*
\end{lstlisting}

In practice, selecting aperture size individually for each object
was found to result only in a minor improvement of photometric accuracy
compared to the use of a properly-selected single-size aperture for all
sources on a frame. This was previously reported by
\cite{2001phot.work...85D}, see also \cite{1999ASPC..172..317M}.

\subsection{PSF-fitting photometry with {\scshape PSFEx}}
\label{sec:psffitting}

In order to enable PSF-fitting you will need to install 
{\scshape SExtractor} (not disabling PSF-fitting support, see 
Section~\ref{sec:remarks}) 
and {\scshape PSFEx} (along with the libraries they depend on) system-wide.
Then you may run
\begin{lstlisting}
./vast -P ../sample_data/*
\end{lstlisting}
where \texttt{'-P'} tells {\scshape VaST} to use PSF-fitting.

While the simple aperture photometry works well out-of-the box for most
CCD images, {\it PSF-fitting photometry will require fine-tuning of PSF
extraction parameters for each new telescope+camera combination.} 
The PSF extraction parameters are set in \texttt{default.psfex} and
described in the {\scshape PSFEx}
documentation\footnote{\url{http://psfex.readthedocs.io}}.
The two most relevant ones are 
\texttt{PSFVAR\_DEGREES} 
and \texttt{PSFVAR\_NSNAP}.
A user should experiment with different values of these parameters and
select the ones that minimize the lightcurve scatter 
for the bright stars in the field (that may be visualized with
\texttt{./find\_candidates}, see 
\ref{sec:restartwork}) for final processing.
This may be a time-consuming process as each
PSF-fitting photometry run needed to construct lightcurves and estimate
their scatter is taken considerably longer than an aperture photometry run
on the same images. Fortunately, it is sufficient to select the best 
PSF extraction parameters once for a given instrument.
%

The source catalogs generated from the input images are cleaned from 
detections that resulted in bad PSF-fit quality. This reduces the number of
false detections due to cosmic ray hits and hot pixels as well as
objects that could not be properly deblended and hence have their photometry
corrupted.

For faint stars PSF-fitting photometry results in considerably smaller
lightcurve scatter compared to the fixed-aperture photometry. However, 
the accuracy of PSF-fitting photometry of bright stars is actually lower
than that of aperture photometry. The reason is that for the bright stars
the dominating source of errors are the residual uncertainties in
reconstructing spatial and temporal variations of the PSF rather than
background noise (as is the case of faint stars). The quality of PSF-fitting
lightcurves can often be considerably improved by applying a few iterations
of the {\scshape SysRem} procedure discussed in
\ref{sec:examplesysrem}.


\subsection{Improving photometric accuracy with {\scshape SysRem}}
\label{sec:examplesysrem}

The {\scshape SysRem} algorithm proposed by \cite{2005MNRAS.356.1466T}
attempts to remove linear effects that, to a various degree, affect many 
lightcurves in a set. The original paper explains the algorithm using
differential extinction as an example, but actually the algorithm is
applicable to systematic effects of any physical origin, as long as many
stars are affected by it.

After constructing lightcurves in the usual way
(\ref{sec:simpleccdsearch}, \ref{sec:psffitting}, \ref{sec:importlightcurves})
run
\begin{lstlisting}
util/sysrem2
\end{lstlisting}
then use 
\begin{lstlisting}
./find_candidates
\end{lstlisting}
to re-compute the variability indices and inspect the results.
Repeat the procedure until there is not further improvement in lightcurve
scatter. Each \texttt{util/sysrem2} removes no more than one systematic
effect (while a real physical effect may be modeled as multiple linear
effects removed by multiple {\scshape SysRem} iterations). Typically 3--6
{\scshape SysRem} iterations are sufficient to considerably improve quality
of CCD lightcurves. One should avoid unnecessary {\scshape SysRem}
iterations, as after all the detectable systematic effects are cleaned from
the dataset, individual variable objects may start to dominate the {\scshape SysRem}
solution and their real variability may be erroneously removed \citep[e.g.]{2013MNRAS.435.3639R}.

\subsection{Variability search with photographic plates}
\label{sec:photoplateexample}

When dealing with digitized photographic images, they typically first
have to be converted from \texttt{TIFF} (the format commonly produced by scanner software) 
to \texttt{FITS}. The conversion can be performed with the tiff2fits 
tool\footnote{\url{ftp://scan.sai.msu.ru/pub/software/tiff2fits/}}:
\begin{lstlisting}
./tiff2fits -i input.tiff output.fits 
\end{lstlisting}
where the \texttt{-i} argument indicates that a positive (white stars on black
sky) should be produced. Information about the observing time should be
added to the \texttt{FITS} header:
\begin{lstlisting}
util/modhead output.fits JD 2442303.54
\end{lstlisting}
After producing a series of images they may be processed with 
{\scshape VaST} (note that a \texttt{default.sex} with customized detection
parameters is needed)
\begin{lstlisting}
cp default.sex.beta_Cas_photoplates default.sex
./vast -o -j ../photographic_fits_images/*
\end{lstlisting}
Here \texttt{default.sex.beta\_Cas\_photoplates} is the example 
{\scshape SExtractor} parameters file for photographic plates,
\texttt{-o} enables the photocurve magnitude calibration function
described in 
Section~\ref{sec:calibinstmag},
\texttt{-j} enables correction for the linear magnitude trend across the image
(Section~\ref{sec:calibinstmag}).

\subsection{Identification of a variable object}
\label{sec:idexample}

If the image field of view is sufficiently large to be automatically
plate-solved with {\scshape Astrometry.net} you may run the automatic star
identification by pressing \texttt{'U'} key in the lightcurve inspection
window or typing in a free terminal
\begin{lstlisting}
util/identify.sh out12345.dat
\end{lstlisting}
where \texttt{out12345.dat} is the file containing {\scshape VaST}-format
lightcurve of the object you are interested in. This will run a series of
scripts that will plate-solve the first image at which the object is
detected, determine its equatorial coordinates and attempt to match it 
external catalogs as described in Section~\ref{sec:astrometry}.

The ability to plate-solve narrow-field images is limited by the index files
available to {\scshape Astrometry.net} code and processing time.
At the time of writing, the plate-solve servers communicating with {\scshape VaST}
are able to solve only images with the field  of view larger than about $30\prime$.
If a narrow field images need to be solved, it is recommended to install {\scshape Astrometry.net}
code locally on the computer running {\scshape VaST} and supply the code
with only with index files corresponding to the field of view of the images.
{\scshape VaST} will automatically detect a local {\scshape Astrometry.net}
installation if its binaries are found in 
\texttt{PATH}. 

\subsection{Importing lightcurves from other software}
\label{sec:importlightcurves}

It is possible to load lightcurves produced by other software into 
{\scshape VaST}. The lightcurves should be in three-column ("JD~mag~err") 
ASCII files. If such lightcurve files are placed in the directory \texttt{../lcfiles} run
\begin{lstlisting}
util/convert/three_column_ascii2vast.sh ../lcfiles/*
./find_candidates
\end{lstlisting}




\subsection{Transient detection}
\label{sec:transients}

An overview of transient search with {\scshape VaST} is presented in
Section~\ref{sec:transientssec}.
You'll need \texttt{libpng} header files installed in your system for this
mode to work. If they were not present the time you installed {\scshape VaST},
you will need to re-compile it by running
\begin{lstlisting}
make
\end{lstlisting}
The following parameters are recommended:
\begin{lstlisting}
./vast -x7 -uf \ 
../transient_detection_test_Ceres/reference_images/* \ 
../transient_detection_test_Ceres/second_epoch_images/*
util/transients/search_for_transients_single_field.sh
\end{lstlisting}
here \texttt{-x7} tells {\scshape VaST} to accept all detections with {\scshape SExtractor}
flags 7 or lower (a transient may well be blended with background stars or
saturated), \texttt{-u} (or \texttt{--UTC}) indicates that the UTC to TT time system
conversion should not be performed 
(Section~\ref{sec:fitsheader}), 
\texttt{-f} tells the program not to start \texttt{./find\_candidates}.
Instead of staring the interactive display, {\scshape VaST} will start a
web browser displaying an \texttt{HTML} page presenting search results
(Figure~\ref{fig:lutetia}). 
If the {\scshape Astrometry.net} code is installed locally
(Section~\ref{sec:astrometry}), 
it is possible to run the transient search
and perform a basic search for known variables (VSX) and asteroids ({\scshape astcheck})
without having internet connection, but for a full investigation of
transient candidates it is highly recommended to use also the online services.
To keep the number of false candidates low, you will likely need to experiment
with source extraction parameters as described in
\ref{sec:finetuningsourceextraction}.

\subsection{Heliocentric correction}
\label{sec:heliocorr}

As the Earth orbits around the Solar system barycenter, it gets closer or
further from distant celestial objects that are not at the ecliptic poles.
The time light from a distant object needs to travel this extra distance
may be as high as 499~seconds for an object at the ecliptic plane observed
from the closest and furthest points of the Earth orbit. This extra time
should be taken into account for accurate timing analysis.

{\scshape VaST} has a tool 
(based on NOVAS\footnote{\url{http://aa.usno.navy.mil/software/novas/novas_info.php}}
library; \citealt{1989AJ.....97.1197K})
that applies heliocentric (center of the Sun) 
correction to a lightcurve given the equatorial J2000 coordinates of 
the observed object:
\begin{lstlisting}
# If the input lightcurve is in TT
util/hjd_input_in_TT out123.dat 12:34:56.7 +12:34:56
# If the input lightcurve is in UTC
util/hjd_input_in_UTC out123.dat 12:34:56.7 +12:34:56
# The output JDs are always expressed in TT
\end{lstlisting}
If timing accuracy of better than 8\,sec is needed, one should consider
using an external software (e.g. {\scshape VARTOOLS} compiled with 
{\scshape SPICE}\footnote{\url{https://naif.jpl.nasa.gov/naif/}} 
support; \citealt{1996P&SS...44...65A}) to compute barycentric correction instead of heliocentric.
The reason why only the simple heliocentric correction is implemented in {\scshape VaST}
instead of barycentric correction is that the heliocentric correction can be computed using 
a few inline constants in the code, while the computation of barycentric correction requires
a complex Solar system model.

\subsection{Re-formatting a lightcurve for publication}
\label{sec:cutelc}

{\scshape VaST} \texttt{out*.dat} lightcurve files include columns with
information about object pixel coordinates, measurement aperture and local
path to the image files that are not needed for a lightcurve attached to
a publication of VSX database submission. {\scshape VaST} includes a tool
that removes the unnecessary columns and sorts the input lightcurve in JD:
\begin{lstlisting}
util/cute_lc out00268.dat > lc_00268.txt
\end{lstlisting}

\subsection{Split multi-extension \texttt{FITS} image}

{\scshape VaST} cannot properly handle multi-extension images -- 
\texttt{FITS} files containing multiple images obtained 
at different epochs of with different CCD chips. 
Multi-extension images are commonly found in the HST and Swift/UVOT archives.
These images should be split into multiple \texttt{FITS} files each
containing only one image before they can be processed with {\scshape VaST}.
A multi-extension file can be split using the tool
\begin{lstlisting}
util/split_multiextension_fits multiextension_image.fits
\end{lstlisting}

\section{Command line options}
\label{sec:vastcommandlineopt}


\texttt{./vast} is the main program that constructs lightcurves from a set of input images.
It accepts a number of command-line options listed below.
\lstset{breaklines=true}
\begin{lstlisting}
  -h or --help       print help message
  -9 or --ds9        use DS9 instead of pgfv to view FITS images
  -f or --nofind     do not run ./find_candidates after constructing lightcurves
  -p or --poly       use linear instead of polynomial magnitude calibration (useful for good quality CCD images)
  -o or --photocurve use formulas (1) and (3) from Bacher et al. (2005, MNRAS, 362, 542) for magnitude calibration. Useful for photographic data
  -P or --PSF        PSF photometry mode with SExtractor and PSFEx
  -r or --norotation assume the input images are not rotated by more than 3 deg. w.r.t. the first (reference) image
  -e or --failsafe   FAILSAFE mode. Only stars detected on the reference frame will be processed
  -u or --UTC        always assume UTC time system, do not perform conversion to TT
  -k or --nojdkeyword  ignore "JD" keyword in FITS image header. Time of observation will be taken from the usual keywords instead
  -a5.0 or --aperture5.0  use fixed aperture (e.g. 5 pixels) in diameter
  -b200 or --matchstarnumber200  use 200 (e.g. 200) reference stars for image matching
  -y3 or --sysrem3  conduct a few (e.g. 3) iterations of SysRem
  -x2 or --maxsextractorflag 2  accept stars with flag <=2 (2 means 'accept blended stars')
  -j or --position_dependent_correction    use position-dependent magnitude correction (recommended for wide-field images)
  -J or --no_position_dependent_correction   DO NOT use position-dependent magnitude correction (recommended for narrow-field images with not too many stars on them)
  -g or --guess_saturation_limit try to guess image saturation limit based on the brightest pixels found in the image
  -G or --no_guess_saturation_limit DO NOT try to guess image saturation limit based on the brightest pixels found in the image
  -1 or --magsizefilter filter-out sources that appear too large or to small for their magnitude (compared to other sources on this image)
  -2 or --nomagsizefilter filter-out sources that appear too large or to small for their magnitude (compared to other sources on this image)
  -3 or --selectbestaperture for each object select measurement aperture that minimized the lightcurve scatter
  -4 or --noerrorsrescale disable photometric error rescaling
  -5 10.0 or --starmatchraius 10.0 use a fixed-radius (in pixels) comparison circle for star matching
  -6 or --notremovebadimages disable automated identification of bad images
\end{lstlisting}
\lstset{breaklines=false}

\section{Log files}
\label{sec:logfiles}

After processing an image series, apart from the lightcurve files {\scshape VaST} will create a number of
log files: 

\texttt{vast\_summary.log} file summarizes the processing results. It
indicates how many images were successfully processed, which image was 
used as the reference (\ref{sec:refimage}) what time system is used for 
JDs in lightcurve files (Section~\ref{sec:fitsheader}).

\texttt{vast\_image\_details.log} contains information about processing of individual images
including the start time and middle of exposure JD derived from the \texttt{FITS}
header, aperture size used for this image, number of detected sources and how many of them are
matched and if the overall image matching and magnitude calibration were successful, 
image rotation angle with respect to the reference image.

\texttt{vast\_command\_line.log} stores all the command-line arguments
specified for the latest {\scshape VaST} run.

\texttt{vast\_lightcurve\_statistics.log} is the table with raw variability index
values computed for all lightcurves that have a sufficient number of points
(Section~\ref{sec:outputgui}).

\texttt{vast\_lightcurve\_statistics\_normalized.log} is the table with variability index values
normalized by their scatter estimated for each object's magnitude.

\texttt{vast\_lightcurve\_statistics\_format.log} describes the format of
the variability index tables.

\section{Comparison with {\scshape AstroImageJ} and {\scshape Muniwin}}
\label{sec:softwaretest}

In order to check the quality of photometry produced by {\scshape VaST} in
aperture and PSF-fitting modes we compare it with two other aperture
photometry packages: {\scshape AstroImageJ} \citep{2017AJ....153...77C} and
{\scshape Muniwin} (Section~\ref{sec:intro}). These two packages
were selected for comparison as they are free and provide a friendly GUI for lightcurve construction.
As the test dataset\footnote{\url{http://scan.sai.msu.ru/vast/Helmos_test}} 
we used a series of $R$-band images of a candidate cataclysmic
variable Gaia16bnz obtained with the 2.3\,m Aristarchos
telescope\footnote{\url{http://helmos.astro.noa.gr/}}
\citep{2010ASPC..424..422G,2016A&A...585A..19B} using 
a VersArray $2048\times2048$ e2v back-illuminated CCD, cooled by liquid nitrogen.
The images cover the sky area of $5.5\times5.5$ arcminutes.
The images were bias-subtracted and flat-fielded in {\scshape VaST} which
was also used to wright the observing time of each image in the commonly accepted 
format using the \texttt{DATE-OBS} and \texttt{EXPTIME} keywords in the \texttt{FITS}
hearer (Section~\ref{sec:fitsheader}).

As the target is one of the few brightest stars on the frame (which may
cause problems for {\scshape VaST} when running with the default parameters), 
the following {\scshape VaST} command line options (\ref{sec:vastcommandlineopt}) 
were used to run the test in the aperture photometry mode:
\begin{lstlisting}[language=vastmarkup]
./vast -a10 -up --noerrorsrescale --magsizefilter \
--notremovebadimages ../Helmos_Gaia16bnz_fixed_date/*\end{lstlisting}
This requires {\scshape VaST} to use a single aperture 10\,pixels in
diameter for all images, do not perform UTC to TT time conversion
(Section~\ref{sec:fitsheader}), use a linear function (\ref{eq:linear}) 
for frame-to-frame magnitude calibration (Section~\ref{sec:calibinstmag}),
enable the magnitude-size filter (Section~\ref{sec:sextraction}) and disable
the bad image filter (as no similar filtering is offered by the other tools).
For the PSF-fitting photometry we used the following command line:
\begin{lstlisting}[language=vastmarkup]
./vast -P -a10 -up --noerrorsrescale --magsizefilter \
--notremovebadimages ../Helmos_Gaia16bnz_fixed_date/*\end{lstlisting}

For {\scshape AstroImageJ} and {\scshape Muniwin} we 
set the measurement aperture diameter to 10\,pixels and the sky annulus inner
and outer diameters to 15 and 35 pixels. We used a single
comparison star URAT1\,697-107802 \citep{2015AJ....150..101Z} assuming 
$R = 13.648$ -- computed from APASS colors using
(\ref{eq:apassrbandconversion}). The same comparison star was used to set
the zero-point of {\scshape VaST} photometry (\ref{sec:examplemanualmagcalib}). The lightcurves
obtained with the different methods agree within 0.005\,mag ($\sigma$) and
0.003\,mag (MAD). A section of the resulting lightcurves is presented in
Figure~\ref{fig:softwaretest}.

The errorbars reported by {\scshape AstroImageJ} and {\scshape Muniwin}
are a factor of three larger than the ones reported by {\scshape VaST}
because they take into account different sources of errors (resulting from
the difference in magnitude calibration technique). The {\scshape VaST} errorbars are
derived from the combination of the background and photon noise
corresponding to the target (and scaled with the magnitude calibration
function described in Section~\ref{sec:calibinstmag}). They neglect the
uncertainty in magnitude zero-point determination as it is derived from all
stars matched between the current and reference frames. The errorbars
reported by {\scshape AstroImageJ} and {\scshape Muniwin} include
the uncertainty in magnitude zero-point determination which in this case
include the photon and background noise from a single comparison star that
is fainter than the target. Normally, the {\scshape VaST} errorbars are
rescaled following the procedure described by
\citep{2009MNRAS.397.1228W,2017MNRAS.468.2189Z}. However, this rescaling procedure
results in overestimating the errors for the brightest stars 
(if there are few bright stars on the frame) and was disabled for this test.

\begin{figure}
 \centering
 \includegraphics[width=0.48\textwidth,clip=true,trim=0cm 0cm 0cm 0cm]{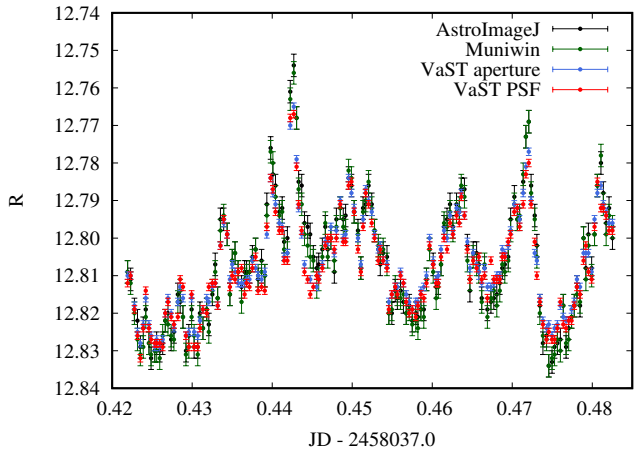}
 \caption{Lightcurve of a candidate cataclysmic variable measured with {\scshape VaST}
 in PSF-fitting and aperture photometry mode, as well as {\scshape AstroImageJ} and 
 {\scshape Muniwin} both performing aperture photometry. }
 \label{fig:softwaretest}
\end{figure}

The candidate cataclysmic variable may be identified as the object
having a higher lightcurve scatter compared to other objects of similar
brightness. Figure~\ref{fig:softwaretestmagdev} presents the lightcurve
scatter versus magnitude plots computed with {\scshape VaST} and 
{\scshape Muniwin}. {\scshape AstroImageJ} has no built-in capability to visualize
magnitude-scatter plots so it is not considered here. {\scshape VaST} and {\scshape Muniwin}
have a different way to quantify lightcurve scatter. In {\scshape VaST} the
default measure of scatter is the unweighted standard deviation computed
over a lightcurve from which 5\,per~cent of brightest and faintest points
were removed, but not more than 5 points from each side (see appendix in \citealt{2017arXiv171007290P}).
In addition to this clipped $\sigma$ {\scshape VaST} has other ways to characterize scatter 
and shape of a lightcurve (Table~\ref{tab:varindex}). {\scshape Muniwin}
uses a custom robust measure of lightcurve scatter as the variability detection statistic.
The few objects with $R\gtrsim16$ showing elevated lightcurve scatter in the {\scshape Muniwin}
plot were automatically identified as blended (Section~\ref{sec:sextraction}) and excluded from the 
{\scshape VaST} analysis.

Figure~\ref{fig:softwaretestmagdev} also illustrates the difference
between the aperture and PSF-fitting photometry with {\scshape VaST}. 
While for the brighter point-like sources both techniques produce 
a comparable measurement accuracy, for fainter sources PSF-fitting results
in much smaller lightcurve scatter than aperture photometry. The effect is
exaggerated here as {\scshape VaST} was forced to use a large aperture 
in order to maximize photometric accuracy for the bright target at
the cost of the elevated background noise mostly affecting the fainter sources.

\begin{figure}
 \centering
 \includegraphics[width=0.48\textwidth,clip=true,trim=0cm 0cm 0cm 0cm]{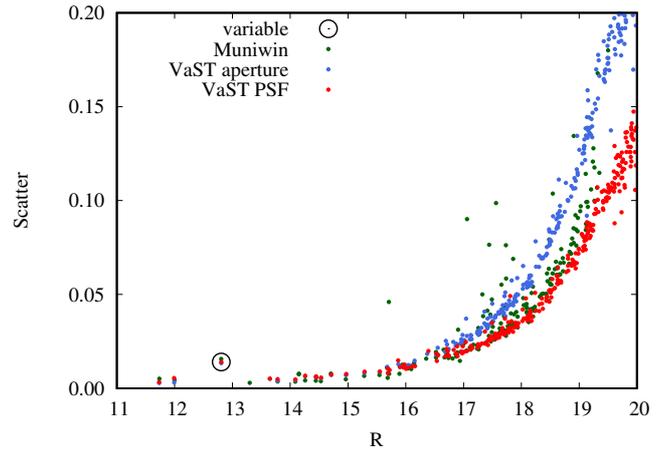}
 \caption{The magnitude-scatter plot produced with {\scshape VaST}
 in PSF-fitting and aperture photometry mode and {\scshape Muniwin}.
 For {\scshape VaST} the Y-axis is the standard deviation computed over
 clipped lightcurves, while for {\scshape Muniwin} the Y-axis
 represents its custom robust measure of lightcurve scatter. The candidate cataclysmic variable
 is marked with the circle. It can be identified as the object with elevated
 lightcurve scatter compared to other objects of similar brightness.}
 \label{fig:softwaretestmagdev}
\end{figure}

\bibliographystyle{model2-names}
\bibliography{vast}







\end{document}